\documentclass{LMCS}

\usepackage{amsmath,amssymb,enumerate}
\usepackage[latin1]{inputenc}
\usepackage{gastex,hyperref}

\newcommand{\intro}[1]{\emph{#1}}

\newcommand{\nats}{\mathbb{N}}
\newcommand{\cbt}{\Delta_2}
\newcommand{\nn}{\Delta_1}
\newcommand{\suc}{\mathit{succ}}

\newcommand{\interpretation}{\mathcal{I}}

\newcommand{\generatorF}{\mathcal{P}^{\scriptstyle W}}
\newcommand{\powerF}{\mathcal{P}^{\scriptstyle F}}
\newcommand{\structure}{\mathcal{S}}
\newcommand{\universe}{\mathcal{U}}
\newcommand{\auto}{\mathcal{A}}

\newcommand{\funind}{\mathit{Index}}
\newcommand{\funsind}{\mathit{SIndex}}
\newcommand{\funbind}{\mathit{BIndex}}
\newcommand{\funrind}{\mathit{RInd}}
\newcommand{\funvind}{\mathit{VInd}}

\newcommand{\set}[1]	{\{{#1}\}}

\newcommand{\kindex}{K_\mathrm{im}}

\newcommand{\ksparse}{K_\mathrm{s}}
\newcommand{\kcomb}{K_\mathrm{c}}

\newcommand{\atoms}{\mathit{Atoms}}
\newcommand{\membership}{\mathit{Mem}}

\newcommand{\dom}{\mathit{dom}}
\newcommand{\qin}{q^{\mathrm{in}}}

\newcommand{\Sig}{\Sigma}
\newcommand{\card}[1]{|#1|}
\newcommand{\st}[1]{N^{#1}_t}

\newcommand{\send}{\mathrm{send}}
\newcommand{\drive}{\mathrm{drive}}
\newcommand{\start}{\mathrm{start}}
\newcommand{\itinerary}{\mathrm{itinerary}}

\newcommand{\phiind}{\phi_{\mathrm{ind}}}
\newcommand{\code}{\mathit{Code}}

\newcommand{\aut}{\text{TA{\scriptsize UT}}}

\newcommand{\cmmt}[1]{}
\newcommand{\dummy}{\diamond}

\newenvironment{theorem}{\begin{thm}}{\end{thm}}
\newenvironment{proposition}{\begin{prop}}{\end{prop}}
\newenvironment{lemma}{\begin{lem}}{\end{lem}}
\newenvironment{corollary}{\begin{cor}}{\end{cor}}
\newenvironment{definition}{\begin{defi}}{\end{defi}}
\newenvironment{remark}{\begin{rem}}{\end{rem}}
\newtheorem{example}[thm]{Example}

\def\doi{3 (2:4) 2007}
\lmcsheading%
{\doi}
{1--36}
{}
{}
{Jul.~18, 2006}
{Mai\phantom{.~0}4, 2007}
{}   

\begin{document}
\title{Transforming structures by set interpretations}

\author[T.~Colcombet]{Thomas Colcombet\rsuper a}
\address{{\lsuper a}Cnrs-Irisa, Irisa, Campus de Beaulieu, 35042
Rennes, France}
\email{Thomas.Colcombet@irisa.fr}

\author[C.~Löding]{Christof Löding\rsuper b}
\address{{\lsuper b}RWTH Aachen, Informatik 7, Ahornstr. 55, Aachen, Germany}
\email{loeding@informatik.rwth-aachen.de}


\begin{abstract}
We consider a new kind of interpretation over relational structures:
finite sets interpretations.  Those interpretations are defined by
weak monadic second-order (WMSO) formulas with free set variables.
They transform a given structure into a structure with a domain
consisting of finite sets of elements of the orignal structure. The
definition of these interpretations directly implies that they send
structures with a decidable WMSO theory to structures with a decidable
first-order theory.  In this paper, we investigate the expressive
power of such interpretations applied to infinite deterministic trees.
The results can be used in the study of automatic and tree-automatic
structures.
\end{abstract}
\maketitle

\section{Introduction}
Computational model theory is concerned with the study of algorithmic
properties of classes of infinite structures
(cf. \cite{BlumensathGra04}), where the focus is on the problem of
model checking such structures against specifications written in some
logic, i.e., deciding for a given structure and logical formula if the
formula holds in this structure. This problem setting has been studied
for various instantiations of the two parameters, i.e., the way to
represent the structures, and the logic to write the
specifications. The most prominent logics in this context are
first-order (FO) logic and monadic second-order (MSO) logic, and they
have led to two tracks of research trying to identify classes of
structures for which the respective logic is decidable.

One way of defining such classes uses, e.g., words or
trees for representing the elements of the structure and uses simple
transformations based on transducers or rewriting to define the
relations of the structure. In this way, one obtains, e.g., the classes
of automatic \cite{hodgson,khoussainovNerode,BlGr00} and
tree-automatic \cite{dauchetTison,BlGr00} structures, for which the
FO-theory is decidable, and the classes of pushdown-graphs
\cite{MullerS85} and prefix recognizable graphs \cite{Caucal96} with
decidable MSO-theory.

In the case of automatic structures, the decidability results are
based on the strong closure properties of finite automata, which are
used to define the relations. Other techniques, e.g. in
\cite{kuskeLohrey} for rewriting in trace monoids, are based on
Gaifman's locality theorem.

The decidability results for MSO logic on pushdown and prefix
recognizable graphs are derived from the results of B\"uchi and
Rabin establishing the equivalence of monadic second-order logic with
certain families of finite automata accepting infinite trees
(cf. \cite{Thomas97}). These results are also underlying the more
recent work \cite{CartonT02} and \cite{unsafe} showing the
decidability of MSO logic over certain classes of infinite words and
infinite trees, respectively.

A different and more systematic approach for defining and studying
classes of infinite structures is to use operations for transforming
structures. An important operation of this kind is the model-theoretic
interpretation. Such an interpretation defines a new structure
`inside' a given one by means of logical formulas describing the
domain and the new relations. Depending on whether these defining
formulas are FO or MSO one speaks of FO- and MSO-interpretations. An
important property of these interpretations is that decidability
results easily transfer from the given structure to the resulting
structure, i.e., applying an FO-interpretation to a structure with a
decidable FO-theory results in a structure with decidable FO-theory,
and similarly for MSO.

As mentioned in \cite{BlumensathGra04}, this suggests a new way of
defining interesting classes of infinite structures: fix an underlying
structure (with good algorithmic properties) and consider all
structures that can be obtained by applying interpretations of a
certain kind. In this way, one obtains the automatic structures by
FO-interpretations from, e.g., a suitable extension of Presburger
arithmetic \cite{blumensath99}, and the prefix recognizable structures
by MSO-interpretations from the infinite binary tree \cite{Bl01}.
This idea has been pursued further in \cite{Caucal02}, where
MSO-interpretations and unravelling of graphs are iterated, leading to
an infinite hierarchy of graphs (or structures) with a decidable
MSO-theory.

All the methods and results described so far can be separated into
those concerned with FO logic (sometimes extended by a reachability
relation \cite{dauchetTison,colcombet02}) and those dealing with MSO
logic (sometimes with only restricted kind of set quantification as in
\cite{Madhusudan03}). To our knowledge, there has been no systematic
work on relating these two areas. In this paper, we bridge this gap by
studying a new kind of interpretation, named finite sets
interpretation, allowing to define classes of structures with
decidable FO-theory from structures with decidable MSO-theory. To be
more precise, we are considering weak MSO (WMSO) logic, i.e., MSO
logic where quantification is restricted to finite sets.  The idea for
these interpretations is rather simple: the domain of the new
structure does not consist of elements of the old structure but of
finite sets of elements of the old structure. The relations are
specified by WMSO-formulas with free set variables (the number of
which corresponds to the arity of the relation). In this way,
FO-formulas over the new structure can directly be translated into
WMSO-formulas over the old structure.  The use of WMSO ensures that
the universe of the resulting structure is countable. It is not clear
whether using standard MSO and then restricting to those resulting
structures with a countable universe gives the same class of
structures.

Using the equivalence of WMSO logic and finite automata (over finite
words and trees) it is not difficult to see that the classes of
automatic and tree automatic structures can be obtained by finite sets
interpretations from $(\nats, \mathit{succ})$, i.e., the natural
numbers with successor relation, and from the infinite binary tree,
respectively (see \cite{colcombetAuckland}). The connection between
automatic structures and finite sets interpretations applied to
$(\nats, \mathit{succ})$ has already appeared before in the
literature. In \cite{ElgotR66} the authors show that the infinite
binary tree together with the equal level relation can be generated
from $(\nats, \mathit{succ})$ by a finite sets interpretation. Today
this extension of the infinite binary tree is known as a generator of
automatic structures in the sense that every automatic structure can
be obtained from it by a first-order interpretation.  A similar result
is given in
\cite{Rubin04} but for another generator of the automatic structures.

This raises the question of what happens when we apply finite sets
interpretations to other structures with decidable WMSO-theory, e.g.,
the structures from the hierarchy defined in \cite{Caucal02}. Though
this hierarchy is strict, it is not a priori clear whether this is
also true for the hierarchy obtained after applying finite sets
interpretations. To answer questions of this kind one has to study
the expressiveness of finite sets interpretations
and to provide tools for showing that a
structure cannot be obtained by such an interpretation applied to a
given structure. In particular, such tools can then be used to answer
questions on automatic structures because these can be obtained by
finite sets interpretations (as mentioned above).
More precisely, we concentrate ourselves in understanding
what are the structures which are finite sets
interpretatable in trees. All the examples mentionned so far
fall in this category.

We contribute to this study via two results.
The first one, Theorem~\ref{tHeorem:injective-presentation},
establishes that the quotient of a structure
finite sets interpretable in a deterministic tree
is itself finite sets interpretable in that tree.
This result was known for automatic structures, and 
was open for tree-automatic structures.
Theorem~\ref{tHeorem:injective-presentation} is a generalisation
of those two cases.

The second and main result,
Theorem~\ref{tHeorem:main-result}, allows to
reduce questions on definability by finite sets interpretations to
questions on WMSO-interpretability.
A precise formulation of it (in its simplest form,
see Corollary~\ref{corollary:main-secondary}) reads as
follows: If the class of structures definable by finite sets
interpretations from a structure $\structure$ is included in the class
of structures definable by finite sets interpretations from a
deterministic tree $t$, then $\structure$ is WMSO-interpretable in~$t$.
Those questions of WMSO-interpretability in trees are well understood
since they can be reformulated in terms of clique-width.
The clique-width is a measure of the complexity of
graphs which has been first introduced for finite graphs \cite{Courcelle89},
and then extended to infinite ones \cite{Courcelle04}. In this latter case,
the equivalence is expressible as follows: ``a graph is WMSO-interpretable
in a tree iff it is of bounded clique-width''.
Our result implies that if we can show that $\structure$ is not
WMSO-interpretable in $t$, then there are structures that can be
obtained by a finite sets interpretation from $\structure$ but not
from $t$. A more technical formulation of the main result also
explicitly gives such a structure.

We demonstrate the use of Theorems~\ref{tHeorem:injective-presentation}
and~\ref{tHeorem:main-result} by showing some non-definability
results, the strictness of the hierarchy mentioned above, and a result
on intrinsic definability of relations related to similar questions
studied for automatic structures (cf. \cite{Barany06}).

The remainder of the paper is structured as follows.  In
Section~\ref{section:definition} we give the basic definitions and
introduce finite sets interpretations, and in
Section~\ref{section:automatic} we give the connection to automatic
structures.  Section~\ref{section:main} is devoted to the study of
finite sets interpretations applied to trees. In particular, our two
results, Theorems~\ref{tHeorem:injective-presentation}
and~\ref{tHeorem:main-result}
are stated in this section. In
Section~\ref{section:applications} we present some applications of our
results. Finally,
Section~\ref{section:result-proof} is devoted to the proof of
the main result.

\section{Definitions and elementary results}\label{section:definition}
In this section we provide the basic definitions used in the paper,
i.e., relational structures, trees, logic, automata, and finally
interpretations, the main subject of this work. We end this section by
giving some elementary results on finite sets interpretations.

\subsection{Structures and trees}
\label{subsection:structures}

We consider (relational) \intro{structures}
$\structure=(\universe,R_1,\dots,R_N)$ where $\universe$ is the
universe of the structure and for each~$i$, $R_i\subseteq
\universe^{r_i}$ is a \intro{relation} of \intro{arity} $r_i$ for a
natural number $r_i$.  The names of the~$R_i$ together with their
arities form the \intro{signature} of the structure.
Trees, as defined below, can be seen in a natural way as particular
instances of such structures.

We will be dealing with infinite binary labeled trees.  From now, we
simply write `trees'. Formally, a tree labeled by a finite
alphabet~$\Sigma$ is a partial mapping $t:\{0,1\}^*\rightarrow \Sigma$
with prefix closed domain $\dom(t)$, and such that if~$u1\in\dom(t)$
then also~$u0\in\dom(t)$.  The elements of the domain are called
\intro{nodes}.  A node~$u$ such that $u0$ is also a node is called an
\intro{inner node}, else it is called a \intro{leaf}.  By
$\sqsubseteq$ we denote the prefix ordering on nodes, also called the
\intro{ancestor} order.  For technical simplifications we will mostly
consider purely binary trees, i.e., such that every node is either a
leaf or has two sons.

Seen as a structure a tree labeled by~$\Sigma$ has as universe the
domain of the tree and contains the following relations: the unary
relations~$S_0$ and~$S_1$ meaning `being a left successor (resp. a
right successor)' (for~$i\in\{0,1\}$, $S_i(u,v)$ holds if~$v=ui$) and
for each $a\in\Sigma$ a unary relation~$a$ interpreted as the set of
elements sent to~$a$ by~$t$.

We will be considering two particular infinite trees,
namely~$\Delta_1$ and~$\Delta_2$.  The tree~$\Delta_1$ is the
unlabeled tree of domain~$0^*$.  We will identify in a natural way
this tree with the structure~$(\nats,\mathit{succ})$.  The
tree~$\Delta_2$ is the unlabeled tree of domain~$\{0,1\}^*$, also
called the \intro{infinite binary tree}.

\subsection{Logic and automata}
\label{subsection:logic}

We use the standard definitions for first-order (FO) and weak monadic
second-order (WMSO) logic. FO-formulas are built up from atomic
formulas using first-order variables (interpreted by elements of the
structure and usually denoted by letters $x$, $y$, $z$) and the
relation symbols from the signature under consideration. Complex
formulas are constructed using boolean connectives and quantification
over first-order variables.

For WMSO-formulas one can additionally use monadic second-order
variables (interpreted by \emph{finite} sets of elements of the
structure and usually denoted by capital letters $X$, $Y$, $Z$),
quantification over such variables, and the membership relation $x \in
X$. If the variables are interpreted by arbitrary sets instead of
finite sets, then we speak of MSO.

In order to deal with WMSO-formulas on trees, we use automata.  Those
automata are more general than WMSO-formulas since they have the
expressiveness of full MSO logic on trees. But for our purpose this
doesn't harm because we only use the translation in one direction,
namely from formulas to automata, and on deterministic trees one can
define finiteness of a set in MSO (a set is finite if its prefix
closure does not contain an infinite path), meaning that for each
WMSO-formula there is an equivalent MSO-formula.

Technically, we use \intro{nondeterministic parity automata} (or
simply \intro{automata}), which are tuples
$(\Sigma,Q,\qin,\delta,\Omega)$ with a finite set $Q$ of
\intro{states}, \intro{initial state} $\qin$, transition relation
$\delta\subseteq Q\times Q\times\Sigma\times Q \uplus \Sigma\times Q$,
and \intro{priority mapping} $\Omega:Q\rightarrow\nats$.  Recall that
we only consider binary trees.  Given a tree~$t$ and an automaton, a
\intro{run} of this automaton on $t$ is a mapping~$\rho: \dom(t)
\rightarrow Q$ such that $(\rho(u0),\rho(u1),t(u),\rho(u))\in\delta$
for each inner node~$u$, and $(t(v),\rho(v))\in\delta$ for every
leaf~$v$.  A run is \intro{accepting} if $\rho(\epsilon)=\qin$ and for
all infinite branches (maximal totally ordered sequences of nodes)
$v_1,v_2,\dots$, $\liminf_{i}\Omega(\rho(v_i))$ is even. We say that a
tree $t$ is accepted by an automaton if there is an accepting run of
this automaton on $t$.  For basic properties of such automata (such as
closure under the Boolean operations) and their relation to logic, we
refer the reader to \cite{Thomas97}.

We are interested here in automata running on a fixed underlying tree
$t$ with additional markings representing (tuples of) subsets of its
domain. To mark a certain subset $X$ of $\dom(t)$ we can put
additional labels on the tree. Formally, the tree $t$ annotated by $X$
is the tree with the same domain as $t$ and labels from $\Sigma \times
\{0,1\}$, where a node $u$ is labeled by the pair $(t(u),0)$ if $u
\notin X$ and $(t(u),1)$ if $u \in X$. In the same way one can also
annotate a tree with tuples of subsets of its domain using a separate
$\{0,1\}$-component for each set.

If $t$ is fixed and $X_1, \ldots, X_n$ are subsets of
$\dom(t)$, then we say that an automaton accepts the tuple $(X_1,
\ldots, X_n)$ if it accepts $t$ with the additional labelings
corresponding to the tuple  $(X_1, \ldots, X_n)$ as explained
above. If we consider all the tuples accepted by an automaton, we
obtain a relation over the subsets of $\dom(t)$. We call this relation
the relation recognized or accepted by the automaton.

A WMSO-formula with free set variables $X_1, \ldots, X_n$ also defines
a relation over the subsets of $\dom(t)$. Throughout the paper we make
use of the following result stating that for each WMSO-formula there
is an equivalent automaton. The proof of this can easily be inferred
from the equivalence of MSO and automata over trees
and from the fact that over trees each
WMSO-formula can be translated into an equivalent MSO-formula.
\begin{theorem}[cf. \cite{Thomas97}]
For each WMSO-formula there is an automaton such that for each tree
$t$ the relation over subsets of $\dom(t)$ defined by the formula is
the same as the one accepted by the automaton.
\end{theorem}

\subsection{Interpretations}
\label{subsection:interpretations}

Interpretations are a standard tool in logic allowing to define
transformations of structures by means of logical formulas.
This technique allows easy transfer of theories from one structure
onto another.
\begin{definition}
An interpretation is a tuple $(\delta,\Phi_{R_1},\dots,\Phi_{R_N})$ of
formulas.  The formula~$\delta$ has only one free variable, and each
formula~$\Phi_{R_i}$ has~$r_i$ free variables.  By our convention, for
weak monadic variables we use capital letters~$X$ and
$X_1,\dots,X_{r_i}$, and small letters~$x$ and $x_1,\dots,x_{r_i}$ in
the case of first-order variables.

An interpretation is FO if the formulas are first-order (and hence the
free variables are also of first-order).  An interpretation is WMSO if
the formulas are weak monadic and the free variables are first-order.  An
interpretation is \intro{finite sets} if the formulas are weak monadic
and the free variables are weak monadic.
\end{definition}
The application of an interpretation to a structure is defined in the
standard way. The only difference is that for finite sets
interpretations the elements of the obtained structure are finite
subsets of the universe of the original structure instead of elements
of the original structure.  Formally, given a structure~$\structure$
and an interpretation $\interpretation=(\delta, \Phi_{R_1}, \dots,
\Phi_{R_N})$, the structure~$\interpretation(\structure)$ has for
universe
\begin{itemize}
\item $\{u \in \universe^\structure~:~\structure\models\delta(u)\}$
	if $\interpretation$ is a FO or WMSO interpretation,
\item $\{U\subseteq\universe^\structure~:~U~\text{finite},~\structure\models\delta(U)\}$
	if $\interpretation$ is a finite sets interpretation,
\end{itemize}
and the interpretation of each symbol~$R_i$ is defined by
\begin{align*}
R_i&=\{(U_1,\dots,U_{r_i})\in(\universe^{\interpretation(\structure)})^{r_i}~:~\structure\models\Phi_i(U_1,\dots,U_{r_i})\}.
\end{align*}
One can note at this point that natural sets interpretations can as
well be defined in a similar way, simply by removing the finiteness
hypothesis on sets and using MSO instead of WMSO.  At the end of this
section we briefly comment on such possible variants for the
definition.

The following example already appears in \cite{ElgotR66}. It shows how
the complete binary tree extended with the equal level relation can be
defined by a finite sets interpretation from the natural numbers with
successor relation. This extension of the binary tree is well-known as
a generator of the automatic structures, in the sense that each
automatic structure can be obtained from it by an FO-interpretation
(cf. \cite{BlGr00}).

%%%%%%%%%%%%%%%%%%%%%%%%%%%%%%%%%%%%%%%% EXAMPLE %%%%%%%%%%%%%%%%%%%%%%%%%%%%%%%%%%%%%%%%
\begin{example} \label{ex:MSO-to-FO}
We show how to obtain the structure $(\{0,1\}^*, S_0, S_1,
\sqsubseteq, \mathit{el})$, i.e., the infinite binary tree extended with the
prefix and equal level relations, by a finite sets interpretation
from the infinite unary tree $\nn$,
i.e., from the natural numbers with
successor $\nn = (\nats,\mathit{succ})$. To realize this we have to
code the nodes of the tree by finite sets of natural numbers and to
describe the relations $S_0$ (for left successor), $S_1$ (for right
successor), $\sqsubseteq$ (for prefix), and $\mathit{el}$ (for equal
level) by means of WMSO formulas.

The coding of the nodes is depicted in Figure~\ref{fig:WMSO-tree}. A
node $u \in \{0,1\}^*$ is represented by the set of positions
corresponding to letter~$1$ in~$u$ and additionally by its length. For example, the node
$100$ is coded by $\{0,3\}$ because its length is $3$ and only
position $0$ is labeled $1$.
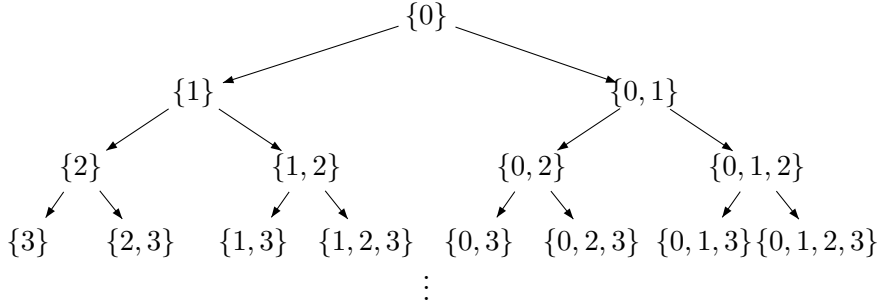
\begin{figure*}
\begin{center}
\begin{picture}(130,50)
\node[Nframe=n](wurzel)(63,40){$\{0\}$}
\node[Nframe=n](0)(32,30){$\{1\}$}
\node[Nframe=n](1)(92,30){$\{0,1\}$}
\node[Nframe=n](00)(17,20){$\{2\}$}
\node[Nframe=n](01)(47,20){$\{1,2\}$}
\node[Nframe=n](10)(77,20){$\{0,2\}$}
\node[Nframe=n](11)(107,20){$\{0,1,2\}$}
\node[Nframe=n](000)(10,10){$\{3\}$}
\node[Nframe=n](001)(25,10){$\{2,3\}$}
\node[Nframe=n](010)(40,10){$\{1,3\}$}
\node[Nframe=n](011)(55,10){$\{1,2,3\}$}
\node[Nframe=n](100)(70,10){$\{0,3\}$}
\node[Nframe=n](101)(85,10){$\{0,2,3\}$}
\node[Nframe=n](110)(100,10){$\{0,1,3\}$}
\node[Nframe=n](111)(115,10){$\{0,1,2,3\}$}
\drawedge(wurzel,0){}
\drawedge(wurzel,1){}
\drawedge(0,00){}
\drawedge(0,01){}
\drawedge(1,10){}
\drawedge(1,11){}
\drawedge(00,000){}
\drawedge(00,001){}
\drawedge(01,010){}
\drawedge(01,011){}
\drawedge(10,100){}
\drawedge(10,101){}
\drawedge(11,110){}
\drawedge(11,111){}
\node[Nframe=n](dots)(63,5){$\vdots$}
\end{picture}
\end{center}
\caption{The nodes of the infinite binary tree $\cbt$ coded by
sets of natural numbers} \label{fig:WMSO-tree}
\end{figure*}
We now define the finite sets interpretation $\interpretation =
(\delta, \Phi_{S_0}, \Phi_{S_1}, \Phi_{\sqsubseteq},
\Phi_{\mathit{el}})$ such that $\interpretation(\nn)$ yields the binary
tree depicted in Figure~\ref{fig:WMSO-tree} together with the
relations $\sqsubseteq$ and $\mathit{el}$. In the formulas we use
abbreviations like $<$ and $\max$ that can easily be defined by
WMSO-formulas in $\nn$.
\begin{itemize}
\item $\delta(X) := \exists x (x \in X)$ (all finite sets except
  $\emptyset$ are used in the coding).
\item $\Phi_{\sqsubseteq}(X_1,X_2) := \max(X_1) < \max(X_2) \land
\forall x (x < \max(X_1) \rightarrow (x \in X_1 \leftrightarrow x \in X_2))$.
\item $\Phi_{S_0}(X_1,X_2) := \Phi_{\sqsubseteq}(X_1,X_2) \land
\max(X_2) = \max(X_1) + 1 \land \max(X_1) \notin X_2$.
\item $\Phi_{S_1}(X_1,X_2):= \Phi_{\sqsubseteq}(X_1,X_2) \land
\max(X_2) = \max(X_1) + 1 \land \max(X_1) \in X_2$.
\item $\Phi_{\mathit{el}}(X_1,X_2) := \max(X_1) = \max(X_2)$
\end{itemize}
\end{example}
%%%%%%%%%%%%%%%%%%%%%%%%%%%%%%%%%%%%%%%% END EXAMPLE %%%%%%%%%%%%%%%%%%%%%%%%%%%%%%%%%%%%%%%%

Let us proceed with some elementary considerations.
Obviously, finite sets interpretations are not closed under
composition. But, as stated in the following proposition, applying an
FO-interpretation after, or a WMSO-interpretation before a
finite sets interpretation, does not give more expressive power.
\begin{proposition}\label{proposition:composition-interpretation}
Let $\interpretation_1$ be a FO-interpretation, $\interpretation_2$
be a WMSO-interpretation, and $\interpretation$ be a
finite sets interpretation. Then $\interpretation_1 \circ
\interpretation \circ \interpretation_2$ is
effectively a finite sets interpretation.
\end{proposition}
\begin{proof}
As for standard interpretations.
\end{proof}
A straightforward as well as essential consequence of this
is expressed in the following corollary.
\begin{corollary} \label{corollary:decidability}
The image of a structure of decidable WMSO-theory by
a finite sets interpretation has a decidable FO-theory.
\end{corollary}

To finish our elementary considerations on finite sets interpretations,
we present Proposition~\ref{proposition:powerset-generator}
which is  a form of converse to Proposition~\ref{proposition:composition-interpretation}.
It states that every finite sets interpretation can be described
as the composition of a specific one, called the weak powerset interpretation,
and a first-order interpretation.
\begin{definition}
Let~$\generatorF$ be the finite sets interpretation
that sends every structure~$\structure$ of signature~$\Sig$
onto a structure of signature~$\Sig\cup\{\preceq\}$
where~$\preceq$ is a new binary symbol, such that
\begin{itemize}
\item the universe is the set of finite subsets of the universe
	of~$\structure$,
\item each symbol~$R$ in~$\Sig$ has the same interpretation as in
	$\structure$ but over singletons instead of elements,
\item the interpretation of~$\preceq$ corresponds to the subset
	ordering.
\end{itemize}
The interpretation $\generatorF$ is called the
\intro{weak powerset} interpretation.
\end{definition}
This interpretation allows to reconstruct all other
finite sets interpretations as stated in the following proposition
which is obtained by a simple syntactic translation of formulas.
\begin{proposition} \label{proposition:powerset-generator}
For each finite sets interpretation~$\interpretation$ there exists a
FO-in\-ter\-pre\-tation~$\interpretation_1$ such that
$\interpretation_1\circ\generatorF = \interpretation$.
\end{proposition}

\subsubsection*{Possible variants}

The definition of finite sets interpretations that we provide uses
WMSO and one might wonder why we restrict ourselves to this logic.
At least two modifications of the definition are very natural:
\begin{itemize}
\item replace WMSO logic with MSO logic in the formulas (but still
only consider finite sets for the free variables),
\item use full MSO and produce a structure over the powerset of the
original structure instead of its finite subsets.
\end{itemize}
If we use the first extension, then
Proposition~\ref{proposition:powerset-generator} fails.  All other
results in this paper would remain unchanged.

The second extension leads to what can be naturally called \intro{sets
interpretation}.  The formulas are MSO, as well as the free
variables.  When applying such a sets interpretation to a structure,
one produces a structure of universe included in the powerset of the
universe of the original structure.

All the results presented above remain valid for this variant.
However, all the results developed below in this work make explicit
use of the finiteness of the sets representing elements of the new
structure.  In particular, one can conjecture that
Theorem~\ref{tHeorem:injective-presentation} would fail for sets
interpretations.  While the conjecture would be that
Theorem~\ref{tHeorem:main-result} still holds for sets
interpretations.

This new kind of interpretation allows to define structures of
uncountable universe from structures of countable universe. This makes
a major difference with respect to finite sets interpretations.  In
general, the relation between set interpretations and finite set
interpretations is not yet understood.  In particular, we do not know
whether the countable structures obtainable by sets interpretations
from trees coincide with the structures obtainable by finite sets
interpretations from trees.

\section{Automatic-like structures} \label{section:automatic}
The line of research that has inspired finite sets interpretations is
the one of automatic structures. Automatic structures in their common
acceptation are structures with a universe consisting of a regular
language of words, and the relations defined by half-synchronized
transducers. The key reason for introducing such structures is that
--- thanks to the good closure properties of finite automata --- they
naturally possess decidable first-order theories.

Classical variants of those structures consider universes consisting
of infinite words ($\omega$-automatic structures), or consisting of
trees, finite or infinite (namely the tree-automatic and
$\omega$-tree-automatic structures).  Some definitions also allow to
quotient the structure by a congruence (that is defined in the same
way as the other relations).  This extension does not increase the
expressiveness of automatic nor tree-automatic structures (cf.
Corollary~\ref{corollary:quotient} below).  The question whether
quotienting increases the expressive power of~$\omega$-automatic and
$\omega$-tree-automatic structures is open.

Historically, automatic as well as $\omega$-automatic structures have
been introduced by Hodgson~\cite{hodgson}.  Khoussainov and Nerode
introduce the notion of automatically presentable theories
\cite{khoussainovNerode}, starting the study of definability in
automatic structures.  The extension to tree-automatic structures can
be traced back, in a different framework, to the work of Dauchet and
Tison \cite{dauchetTison}.  Blumensath and
Grädel~\cite{blumensath99,BlGr00} then formalize the notion of
tree-automatic structure and add to it the family of
$\omega$-tree-automatic structures.  Independently, the study of
3-manyfolds lead to the particular case of automatic
groups~\cite{epstein}.

\subsection{Word-automatic structures}
\label{subsection:wordautomatic}

In the case of words a relation $R \subseteq (\Sigma^*)^r$ is
automatic if there is a finite automaton accepting exactly the tuples
$(w_1, \ldots, w_r) \in R$, where the automaton reads all the words in
parallel with the shorter words padded with a dummy symbol
$\dummy$. Formally, for $w_1, \ldots, w_r \in \Sigma^*$ we define
\newcommand{\tuple}[2]{\left[\begin{array}{c} #1 \\ \vdots \\
#2\end{array}\right]}
\[
w_1 \otimes \cdots \otimes w_r = \tuple{a_{11}'}{a_{r1}'} \cdots
\tuple{a_{1n}'}{a_{rn}'} \in (\Sigma_{\dummy}^r)^*
\]
where $\Sigma_{\dummy} = \Sigma \cup \{\dummy\}$, $n$ is the maximal
length of one of the words $w_i$, and $a_{ij}$ is the $j$th letter of
$w_i$ if $j \le |w_i|$ and $\dummy$ otherwise. A language $L \subseteq
((\Sigma \cup \{\dummy\})^r)^*$ defines a relation $R_L \subseteq
(\Sigma^*)^r$ in the obvious way: $(w_1, \ldots, w_r) \in R_L$ iff
$w_1 \otimes \cdots \otimes w_r \in L$.  A tuple $(L,L_1, \ldots,
L_n)$ of languages $L \subseteq \Sigma^*$ and $L_i \subseteq
(\Sigma_{\dummy}^{r_i})^*$ defines a structure of universe $L$ with
the relations $R_{L_i}$ of arity $r_i$. A structure $\structure =
(\universe, R_1, \ldots, R_n)$ is automatic if it is isomorphic to a
structure of the above kind for regular languages $L,L_1, \ldots,
L_n$.

The class of $\omega$-automatic structures is defined in the same way
with infinite words instead of finite ones. In this case the
definition is even simpler as there is no need for padding shorter
words.

\subsection{Tree-automatic structures}
\label{subsection:tree-automatic}

To define tree-automatic structures we need a way to code tuples of
finite trees, i.e., we need an operation $\otimes$ for finite
trees. For a tree $t:\dom(t) \rightarrow \Sigma$ let $t^{\dummy}:
\{0,1\}^* \rightarrow \Sigma_{\dummy}$ be defined by $t^{\dummy}(u) =
t(u)$ if $u \in \dom(t)$, and $t^{\dummy}(u) = \dummy$ otherwise.  For
finite $\Sigma$-labeled trees $t_1, \ldots, t_r$ we define the
$\Sigma_{\dummy}^r$-labeled tree $t = t_1 \otimes \cdots \otimes t_r$
by $\dom(t) = \dom(t_1) \cup \cdots \cup \dom(t_r)$ and $t(u) =
(t_1^{\dummy}(u), \ldots, t_r^{\dummy}(u))$. When viewing words as
unary trees, this definition corresponds to the operation $\otimes$ as
defined for words.  As in the case of words a set $T$ of finite
$\Sigma_{\dummy}^r$-labeled trees defines the relation $R_T$ by $(t_1,
\ldots, t_r) \in R_T$ iff $t_1 \otimes \cdots \otimes t_r \in T$. A
structure is called tree-automatic if it is isomorphic to a structure
given by a tuple $(T,T_1, \ldots, T_n)$ of regular tree languages in
the same way as for words.

Again, the definitions for $\omega$-tree-automatic structures are the
same with $\omega$-trees, i.e., trees of domain $\{0,1\}^*$ instead of
finite trees.

One should note here that we only consider so called injective
presentations of automatic structures. A more general definition as,
e.g., in \cite{blumensath99} additionally uses a regular language
$L_{\sim} \subseteq (\Sigma_{\dummy}^2)^*$ defining an equivalence
relation identifying words representing the same element of the
structure (and similarly for the other variants of automatic
structures). It is known that injective presentations are sufficient
for automatic structures \cite{khoussainovNerode} meaning that all
structures that are automatic in the more general sense are also
automatic according to our definition.  The corresponding result for
tree-automatic structures is established below, see
Corollary~\ref{corollary:quotient}.

\subsection{Automaticity via interpretations}

Recall that $\nn$ is the (unlabeled) infinite unary tree, i.e., the
natural numbers with successor, and that $\cbt$ is
the (unlabeled) infinite binary tree.
The following fact is a straightforward consequence
of the definition of automatic structures and of the equivalences
between WMSO-logic and automata. The first claim also appears in
\cite[Thm. C.2.11, page 50]{Rubin04}.
\begin{proposition} \label{proposition:automatic}
The following holds up to isomorphism
\begin{itemize}
\item A structure is \intro{automatic} iff it is
finite sets interpretable in~$\nn$.
\item A structure is \intro{tree-automatic} iff it is
finite sets interpretable in~$\cbt$.
\end{itemize}
\end{proposition}
\begin{proof}[Proof (sketch).]
As already mentioned, the first claim is shown in \cite{Rubin04}. 

We describe here how to proceed for tree-automatic structures,
starting by an explanation how to obtain a tree-automatic presentation
from a finite sets interpretation $\interpretation$ in~$\cbt$. A
finite subset $X$ of $\cbt$ is coded as a finite tree as follows: the
tree has the smallest domain that contains all elements from $X$ and
is labeled $0$ at positions that are not in $X$ and $1$ at positions
that are in $X$.  For each formula of $\interpretation$ there is an
equivalent parity automaton over $\cbt$. This automaton can easily be
turned into an automaton over finite trees accepting the corresponding
relation over the codings as just described.

For the other direction we start from a tree-automatic presentation of
a structure. A first thing to note is that a singleton alphabet of tree
labels is sufficient because each symbol from a larger alphabet can be
encoded in a finite pattern; in this construction,
we use the leaf/non-leaf nature of each node for coding information.
Since now the alphabet is a singleton, one naturally encodes a tree~$t$
by the finite set~$\dom(t)$.
Now pick an automaton from the tree-automatic presentation
accepting a relation and pick a tuple of such sets coding finite
trees.  Using standard techniques for translating automata to logic
(cf. \cite{Thomas97}), one describes in WMSO that
the corresponding tuple of finite trees is accepted by the automaton.
\end{proof}

Note that the $\omega$-automatic and $\omega$-tree-automatic structures
satisfy the same equivalences where finite sets interpretations are
replaced by sets interpretations.

\section{Finite sets interpretations on trees}\label{section:main}
The power of finite sets interpretations makes it difficult to obtain
results for the general case where such interpretations are applied to
arbitrary structures.  For this reason, in this article we only
consider the special case of finite sets interpretations applied to
deterministic trees.

This restriction can be justified in two ways.  The first
justification is that on trees there are specific tools suitable for
treating WMSO questions: their automata equivalents.  The second
justification is that if Seese's conjecture \cite{Seese91} holds ---
stating that all structures of decidable weak monadic second-order
theory are WMSO-interpretations of trees --- then the only structures
that we can prove to have decidable first-order theory using
Corollary~\ref{corollary:decidability} are finite sets interpretations
of trees (for recent work on Seese's conjecture see
\cite{CourcelleOum}).

We give here two results concerning finite sets interpretations
applied to deterministic trees.  The first one --- in
Subsection~\ref{subsection:quotient} --- shows that finite sets
interpretations on deterministic trees followed by a quotient are
simply finite sets interpretations.  The second result --- subject of
Subsection~\ref{subsection:main-result} --- concerns finite sets
interpretations applied to trees leading to powerset lattices. The
technical core of the proof is given in
Section~\ref{section:result-proof}.

\subsection{Quotienting finite sets interpretations on trees}
\label{subsection:quotient}

We show here that if a structure is finite sets
interpretable in a deterministic tree containing a symbol
interpreted as a congruence on the structure, then it is
possible to directly obtain the quotiented structure by
a finite sets interpretation.

A \intro{congruence} on a structure~$\structure$ is an equivalence
relation~$\sim$ such that for every symbol~$R$ of arity~$n$ and all
elements $x_1,\ldots, x_n, y_1, \ldots, y_n$ of $\structure$:
if~$x_1\sim y_1,\dots, x_n\sim y_n$, then
$R^\structure(x_1,\dots,x_n)$ iff $R^\structure(y_1,\dots,y_n)$. We
say that a symbol is a congruence if its interpretation in the
structure is a congruence.  For a congruence~$\sim$ over a
structure~$\structure$, we denote by~$\structure/_\sim$ the
\intro{quotient structure}, i.e., the structure which has as elements the
equivalence classes of~$\sim$, and the relations of which are the
images of the relations on~$\structure$ under the canonical surjection
induced by~$\sim$.

Note that the operation of quotienting preserves the decidability of
the first-order theory. For this reason we may wonder if constructing
a structure by a finite sets interpretation followed by a quotient is
more powerful than solely using a finite sets
interpretation. Theorem~\ref{tHeorem:injective-presentation} below
shows that it is not the case when the original structure is a
deterministic tree.

\begin{theorem}\label{tHeorem:injective-presentation}
Given a finite sets interpretation~$\interpretation$, there exists a
finite sets interpretation~$\interpretation'$ such that for every
deterministic tree~$t$,
if the symbol~$\sim$ is a congruence in~$\interpretation(t)$,
then
$\interpretation'(t)$ is isomorphic to $\interpretation(t)/_{\sim}$.
\end{theorem}
\begin{proof}
Let~$\auto$ be a nondeterministic parity automaton corresponding to
the formula $\Phi_{\sim}$ describing~$\sim$ in~$\interpretation$.
This automaton works on $t$ additionally annotated by a pair $(X,Y)$
of sets of nodes. We say that the automaton reads $(X,Y)$.

For a prefix closed subset~$S$ of~$\dom(t)$, we call the set of
minimal nodes not in~$S$ its \intro{border}.  Let~$X$ be an element
of the structure~$\interpretation(t)$, i.e., a finite set of nodes
of~$t$.  We construct its \intro{shadow}~$S(X)$ by taking the prefix
closure of $X$ and then adding all trees of height at most $2^{|Q|}-1$
that are rooted at the border of this prefix closure.
So $S(X)$ is the least set of
nodes containing $X$ that is closed by prefix and such that all element
of its border is the root of a subtree of height at least~$2^{|Q|}$. Obviously, $S(X)$ is
finite.  Now, let us consider an equivalence class $c$ for~$\sim$.
Define the \intro{shadow}~$S(c)$ of the class~$c$ to be the
intersection of the shadows of all the elements in the class.  This is
also a finite set of nodes of~$t$.  We define the border of the class,
denoted by $B(c)$, to be the border of its shadow. Note that the
subtrees rooted at nodes from $B(c)$ have height at least~$2^{|Q|}$;
indeed,
this property is preserved when intersecting shadows.

Given an element~$X$, its description is the triple~$(B,Y,f)$,
where~$B = B(c)$ for the class $c$ of~$X$, $Y$ is~$X \cap S(c)$,
i.e., $X$ restricted to the shadow of its class, and~$f$ maps each
node~$x \in B$ to the set of states~$q$ such that there is an
accepting run of~$\auto$ on the subtree rooted in~$x$ starting with
state~$q$ and reading $(\emptyset,X)$.

We claim that if two elements~$X$ and~$X'$ share the same description
--- say $(B,Y,f)$ --- then those elements are equivalent for~$\sim$.
Using the transitivity of~$\sim$ and the finiteness of~$B$ it is
sufficient to consider the case of elements coinciding everywhere but
below one~$x$ in~$B$. Note that $X$ and $X'$ coincide above $B$
because they have the same description.  Since~$x$ is in the border
of the class of~$X$, there is an element~$Z$ equivalent to~$X$ such
that~$Z$ does not contain any node below~$x$.  Since~$Z$ is equivalent
to~$X$, there is an accepting run~$\rho$ of~$\auto$ on~$t$ labeled
by~$(Z,X)$.  We aim at constructing a run of~$\auto$ witnessing the
equivalence of~$Z$ and~$X'$, i.e., a run accepting $t$ labeled
by~$(Z,X')$. This new run is constructed in the following way. On
every element not below~$x$, it coincides with~$\rho$. This is a valid
part of run since~$X$ and $X'$ do coincide on this area.  On the
subtree rooted in~$x$, $Z$ coincides with~$\emptyset$.  Hence,
as~$(B,Y,f)$ is a description of~$X$, $\rho(x)$ belongs to~$f(x)$. But
as the same description holds also for~$X'$, there is a piece of run
below~$x$ starting with~$\rho(x)$ and accepting~$(\emptyset,X')$. We
complete our new run by this piece of run.  This new run witnesses as
expected that~$Z\sim X'$. It follows by symmetry and transitivity
of~$\sim$ that~$X\sim X'$.  This concludes the proof of the claim.

Let us remark now that a description~$(B,Y,f)$ can be encoded uniquely
by a set of nodes: this set is~$B\cup Y\cup \mathit{Coding}(f)$
where~$\mathit{Coding(f)}$ contains exactly one element for each
element~$x$ of the border, and this element is located in a place
uniquely describing the value of~$f(x)$ (e.g. the leftmost node at
distance $g(f(x))$ below~$x$,
where~$g$ is a numbering of~$2^Q$ starting from~$1$).  This is
possible since the subtree rooted in~$x$ is has height at least~$2^{|Q|}$,
and consequently, there is ``room'' below~$x$ for coding the
information~$f(x)$.

Note that associating to an element~$X$ the coding of its description
is doable by means of a WMSO-formula.  Note also that given a
class~$c$ there is only a finite number of descriptions for the
elements it contains.  Hence we can choose the smallest description
--- smallest for a suitable total order --- as unique representative
for the class. A suitable total order can, e.g., use the
lexicographically smallest node that is in the symmetric difference of
the two sets coding the two descriptions.  From here, it is not
difficult to reconstruct~$\interpretation(t)/_\sim$\ .
\end{proof}

In combination with Proposition~\ref{proposition:automatic} we obtain the following.
\begin{corollary}\label{corollary:quotient}
Tree-automatic structures are effectively closed under quotient.
\end{corollary}
In the terminology of~\cite{blumensath99}, this result is rephrased as
``every tree-automatic structure admits an injective presentation.''
Let us remark that this result is announced
in~\cite{blumensath99}, but unfortunately the proof proposed there
contains an unrecoverable error.

\subsection{Finite sets interpretations and powerset lattices}
\label{subsection:main-result}

In this section we present our main result.  For this,
let~$\structure$ be a structure of signature $\preceq$. We say
that~$\structure$ is a \intro{finite powerset lattice} if it is
isomorphic to~$(\powerF(E),\subseteq)$ for some set~$E$,
where~$\powerF(E)$ represents the finite subsets of~$E$.  Such a
finite powerset lattice can be seen as a particular case of weak
powerset generator applied to a vocabulary-free structure.  We call
the elements corresponding to singletons in this isomorphic structure
\intro{atoms}, i.e., those elements which have exactly one element
strictly smaller with respect to~$\preceq$.
\begin{theorem}\label{tHeorem:main-result}
For every finite sets
interpretation~$\interpretation=(\delta(X),\phi_\preceq(X,Y))$,
there exists a WMSO-formula $\code(X,x)$ such that, whenever for some
tree~$t$, $\interpretation(t)$ is a finite powerset lattice, then
$\code(X,x)$ evaluates on~$t$ to  an injection mapping the atoms
of~$\interpretation(t)$ to nodes of~$t$.
\end{theorem}
Section~\ref{section:result-proof} is dedicated to the rather long and
involved proof of this result.

We rarely use the theorem in this form.  We rather use weakened
versions of it, namely Corollary~\ref{corollary:main-general} and
Corollary~\ref{corollary:main-secondary}.
\begin{corollary}\label{corollary:main-general}
For every finite sets interpretation~$\interpretation$,
there exists a WMSO-in\-ter\-pre\-ta\-tion~$\interpretation_2$
such that whenever for some structure~$\structure$
and some tree~$t$, $\interpretation(t)$ is
isomorphic to~$\generatorF(\structure)$
then $\interpretation_2(t)$ is isomorphic to~$\structure$.
\end{corollary}
\begin{proof}
If we remove all relations other than~$\preceq$,
the weak powerset generator is nothing but a finite powerset
lattice. Hence we can obtain a formula~$\code(X,x)$
by application of Theorem~\ref{tHeorem:main-result}.
It is then easy to transfer all relations defined
on singletons to their image by~$\code$.

Formally, we define the WMSO-interpretation $\interpretation_2 =
(\delta, \Phi_{R_1}, \dots, \Phi_{R_l})$ as follows:
\begin{itemize}
\item $\delta(x)=\exists X.\code(X,x)$,
\item for each symbol~$R$ of arity $r$ from the signature,
$\Phi_R(x_1,\dots,x_{r})$ is defined as
\[
\exists X_1,\dots,X_{r}.\Psi_R(X_1,\dots,X_r) \wedge
\bigwedge_{i=1}^{r} \code(X_i,x_i)
\]
where each~$\Psi_R$ is the WMSO-formula in~$\interpretation$ defining
the interpretation of the symbol~$R$.
\end{itemize}
As $\code$ maps each atom of $\interpretation(t)$ to a unique node,
this interpretation indeed maps $t$ to a structure isomorphic to
$\structure$.
\end{proof}

A weaker, yet more readable formulation
of the above corollary is provided in the following one.
\begin{corollary}\label{corollary:main-secondary}
If for a structure $\structure$ and a tree $t$
the class of structures that can be obtained by
finite sets interpretations from $\structure$ is
contained in the class of structures that can be
obtained by finite sets
interpretations from $t$, then $\structure$
is WMSO-interpretable in~$t$.
\end{corollary}
\begin{proof}
Assume that the hypothesis of the corollary holds.  In particular, the
structure~$\generatorF(\structure)$ is isomorphic
to~$\interpretation(t)$ for some finite sets
interpretation~$\interpretation$.  By applying
Corollary~\ref{corollary:main-general}, the structure~$\structure$ is
WMSO-interpretable in~$t$.
\end{proof}

\section{Applications}\label{section:applications}
We present here several applications of the results above, ordered by
level of complexity.  The two first ones, showing that the free monoid
is not obtainable by a finite sets interpretation of a tree
(Section~\ref{subsection:free-monoid}) and that a natural hierarchy of
structures is strict (Section~\ref{subsection:hierarchy-separation}),
are paradigmatic applications of Theorem~\ref{tHeorem:main-result}.
Section~\ref{subsection:random-graph} establishes that the random
graph is not finite sets interpretable in a tree, extending the known
result for automatic structures \cite{khoussainovNRS04}.  Finally, in
Section~\ref{subsection:intrinsic-regularity}, we study intrinsically
definable relations in generators of the automatic structures.

Some of those results are known in the weaker context of automatic
structures.  We would like to underline here the fundamental
methodological difference of our approach: in none of the applications
below we perform a combinatorial analysis of finite sets
interpretations.  Instead, we systematically reduce the problem to a
much easier one of WMSO-definability in trees.

\subsection{The free monoid}
\label{subsection:free-monoid}
We consider the free monoid as a structure $(\{a,b\}^*, \cdot, a, b)$ ---
the set of words over~$\{a,b\}$ together with the ternary
relation corresponding to the concatenation and the two words $a$
and~$b$ identified by unary predicates --- and want to answer the
question whether this structure is isomorphic to a finite sets
interpretation of a tree.

One should note that the FO theory of the free monoid is undecidable
and hence we can directly conclude that it cannot be obtained by a
finite sets interpretation from a tree with a decidable WMSO
theory. However, this reasoning does not include trees with an
undecidable WMSO theory.

The negative answer we give here to the above question is the simplest
and in some sense the purest application of the results presented
above and should be considered for this reason as a key example.

The following result was obtained in a discussion with Olivier Ly.

\begin{proposition} \label{proposition:free-monoid}
The free monoid over a two letter alphabet is not isomorphic
to any finite sets interpretation of a tree.
\end{proposition}
\begin{proof}
We first show how to obtain $\generatorF(\nats,+)$ from the free
monoid by an FO-interpretation followed by a quotient. Then, assuming
that our claim is false, we invoke the two results from the previous
section and obtain a contradiction.

Let~$f$ be the function which to each word of the form
$ba^{n_1}ba^{n_2}b\dots b a^{n_k}b$ over~$\{a,b\}$ associates the set
of naturals~$\{n_1,n_2,\dots,n_k\}$.  The domain of~$f$ is the set of
elements satisfying $\mathit{dom}(x)=\exists y.x=byb$.  The relation
of inclusion is also first-order definable: $f(u)\subseteq f(v)$ iff
$\mathit{sub}(u,v)$ holds with $\mathit{sub}(u,v)=$
\begin{gather*}
\forall x\in a^*. \exists y,z.\  u=ybxbz \rightarrow \exists y',z'. v=y'bxbz'\ ,\\
\text{where}~x\in a^*~~\text{stands for}~~\forall~y,z. x\neq ybz\ .
\end{gather*}
Let~$x\sim y$ be the formula~$\mathit{sub}(x,y)\wedge\mathit{sub}(y,x)$.
Naturally, for every~$u,v$, $u\sim v$ iff
$f(u)=f(v)$.
Finally the addition over singletons is definable.
More precisely, $f(u)=\{i\}$, $f(v)=\{j\}$ and~$f(w)=\{i+j\}$
iff $add(u,v,w)$ holds with $add(u,v,w)=$
\begin{align*}
\exists x\in a^*,y\in a^*.u\sim bxb\wedge v\sim byb\wedge w\sim bxyb.
\end{align*}
Using those formulas,
one can first-order interpret in the free monoid
a structure which, when quotiented by~$\sim$,
is isomorphic to $\generatorF(\nats,+)$.

Assume now that the free monoid can be finite sets interpreted in some
tree~$t$. Since structures obtainable by finite sets interpretations
from~$t$ are closed under first-order interpretations
(Proposition~\ref{proposition:composition-interpretation}) and
quotient (Theorem~\ref{tHeorem:injective-presentation}), this implies
that $\generatorF(\nats,+)$ is finite sets interpretable in~$t$.  By
Corollary~\ref{corollary:main-general} we deduce that~$(\nats,+)$ is
WMSO-interpretable in~$t$. This yields a contradiction
since~$(\nats,+)$ is not WMSO-interpretable in a tree.
Let us provide a direct argument for proving this last argument.

Assume that $(\nats,+)$ is WMSO-interpretable in $t$ and let $U$
denote the set of nodes of $t$ that represent $\nats$ in a
corresponding interpretation. We first note that for each node of $t$
at most one of its subtrees contains infinitely many elements from
$U$. Otherwise, if the two subtrees of a node $u$ contain both
infinitely many elements from $U$, the successor relation on $\nats$
(which is addition with one argument fixed to $1$) would infinitely
often jump between these two subtrees. If $\auto$ is an automaton with
$n$ transitions accepting the successor relation, and if $x_0, \ldots, x_n
\in U$ are in the left subtree of $u$, $y_0, \ldots, y_n \in U$ are in
the right subtree of $u$, and all $x_i,y_i$ are in the successor
relation, then $\auto$ also accepts a pair $x_i,y_j$ with $i \not=j$
by a simple counting argument on the transitions used at $u$ in
accepting runs of $\auto$.

By starting at the root and always proceeding to the unique subtree
containing infinitely many elements from $U$ we obtain an infinite
branch $B$ of $t$.

Now let $\auto_+$ be an automaton with $n$ transitions accepting the
relation $+$ on $t$ and let $x_0, \ldots, x_n \in U$. For each $x_i$
there are infinitely many $y_i,z_i$ such that the triple
$(x_i,y_i,z_i)$ is accepted by $\auto_+$. Choose a node $v$ on $B$
such that none of the $x_i$ is below $v$ and for each $i$ choose
$y_i,z_i$ as above that are below $v$. Counting the possible
transitions that are used at $v$ in accepting runs of $\auto_+$ on the
tuples $(x_i,y_i,z_i)$ we obtain that $\auto_+$ also accepts
$(x_i,y_j,z_j)$ for some $i \not= j$. This gives a contradiction.
\end{proof}

\subsection{A hierarchy of structures of decidable first-order theory}
\label{subsection:hierarchy-separation}

Caucal~\cite{Caucal02} introduces a hierarchy of graphs/structures of
decidable MSO-theory. The definition of this hierarchy that we use
here differs from the original one and follows \cite{CarayolW03} and
\cite{Thomas03}.

Level~$0$ consists of finite structures, and
level~$n+1$ is defined as the MSO-interpretations of the unraveling of
graphs of level~$n$. As both MSO-interpretation and unraveling are
transformations preserving the decidability of the MSO-theory, each
structure of this hierarchy has a decidable MSO-theory.
In~\cite{CarayolW03}, this hierarchy is shown to be strict.  

If in these definitions the MSO-interpretations are replaced by
WMSO-interpretations, we obtain the same hierarchy. This can be
deduced from a result in \cite{CarayolW03} stating that each graph of
level~$n$ can be obtained from the unravelling of a deterministic
graph of level~$n-1$ by applying a so called inverse rational
mapping. Such an inverse rational mapping can easily be described by a
WMSO-interpretation. 

Furthermore, the unravelling of a deterministic graph yields a
deterministic tree and on deterministic trees every WMSO-formula is
equivalent to an MSO-formula. Hence, from the decidability of the MSO
theory we also obtain the decidability of the WMSO theory for
deterministic trees. In combination with the above mentioned
result all graphs in the hierarchy have a decidable WMSO theory.

From this hierarchy, it is easy to construct a corresponding
tree-automatic hierarchy. The tree-automatic structures of level~$n$
are the image of the structures of level~$n$ of the Caucal hierarchy
by finite sets interpretations. Let us denote the $n$th level of
this tree-automatic hierarchy by $\aut_n$.  Since the trees on the
first level of the Caucal hierarchy are regular, we can deduce from
Proposition~\ref{proposition:automatic} that $\aut_1$ coincide
with the class of tree-automatic structures.

Furthermore, from Corollary~\ref{corollary:decidability} and the above
considerations we can conclude the following decidability result.

\begin{remark}
For each $n$, every structure in $\aut_n$ has a decidable FO theory.
\end{remark}

A simple application of our result is the strictness of this
tree-automatic hierarchy.
\begin{theorem}
For each $n \ge 0$ the class $\aut_n$ of structures on the $n$th level
of the automatic hierarchy is strictly contained in $\aut_{n+1}$.
\end{theorem}
\begin{proof}
We know that each level~$n$ of the Caucal hierarchy contains a tree
generator~$G_n$, i.e., each structure of level~$n$ is
WMSO-interpretable in~$G_n$ \cite{CarayolW03}. Let us suppose that the
automatic hierarchy collapses at some level~$n$, i.e., $\aut_n =
\aut_{n+1}$. This would imply that the
structure~$\generatorF(G_{n+1})$ can be obtained by
a finite sets interpretation from $G_n$. Then, by
Corollary~\ref{corollary:main-secondary}, we obtain $G_{n+1}$ as a
WMSO-interpretation of $G_n$ and hence $G_{n+1}$ is in the $n$th level
of the hierarchy. This contradicts the strictness of the Caucal
hierarchy.
\end{proof}

\subsection{Random graph}
\label{subsection:random-graph}

The \intro{random graph} is a non-oriented unlabeled countable graph
with the following fundamental property: for any two disjoint
finite set of vertices~$E$ and~$F$, there exists a vertex~$v$ that is
connected to all the elements of~$E$ and to none of the elements
of~$F$.  For the existence and basic properties of such a graph see,
e.g., \cite{Hodges93}.  We do not give in this work a more precise
definition of the random graph.  Anyhow, a direct consequence of the
fundamental property stated above is that the random graph satisfies
the quantifier elimination property (and this is effective).
The decidability of its first-order theory follows.

Since the random graph has a decidable first-order theory and since
finite sets interpretations define a large number of structures also
having this property, it is interesting to consider the question
whether the random graph can be obtained by a finite sets
interpretation from a tree.  A partial answer to this question has
been studied: one knows that the random graph is not isomorphic to any
word-automatic structure \cite{khoussainovNRS04}.

In this section, we show that there is no tree from which the random
graph can be generated by a finite sets interpretation. This proof
was obtained in a discussion with Vince B\'{a}r\'{a}ny.

\begin{theorem}\label{tHeorem:random}
The random graph is not finite sets interpretable in a tree.
\end{theorem}
\begin{proof}
Heading for contradiction, let us assume that there exists a
finite sets interpretation~$\interpretation_R=(\delta(X),\Psi(X,Y))$
and a binary tree~$t_R$ such that~$\interpretation_R(t_R)$ is
(isomorphic to) the random graph.  

The basic idea is to prove that, under this assumption,
``the random graph is WMSO-interpretable in a tree''.
However, this statement is uncomfortable to handle since the properties of the random graph do only refer to its finite induced subgraphs.
Instead, we show a similar interpretability result
for every finite induced
subgraph of the random graph; i.e. for every finite graph.
Formaly we establish the following claim.
\par \medskip \par \noindent
\textit{Claim:}
There exists a finite sets interpretation~$\interpretation'$ such that
for any finite non-oriented graph~$G$ there exists a tree~$t_G$ such
that~$\interpretation'(t_G)$ is isomorphic to~$\generatorF(G)$.
\par \medskip \par \noindent
Before we prove this claim, let us demonstrate how to use it to show
Theorem~\ref{tHeorem:random}.  Let us apply
Corollary~\ref{corollary:main-general} on the
interpretation~$\interpretation'$.  We obtain a WMSO-interpretation
$\interpretation''$ with the property that $\interpretation''(t_G)
\cong G$ for any non-oriented unlabeled graph~$G$ and a suitably
chosen tree~$t_G$.  As trees have bounded clique-width and
WMSO-interpretations applied to a class of graphs of bounded
clique-width yield also a class of graphs of bounded clique-width (see
e.g. \cite{Courcelle97}\footnote{To be precise the result in
\cite{Courcelle97} applies to classes of finite graphs, whereas our
trees $t_G$ may be infinite. But given an interpretation that produces
a class of finite graphs from a class of infinite trees one can modify
the interpretation such that one obtains the same resulting class of
graphs from a class of finite trees.}), we obtain a contradiction to
the fact that there exists non-oriented graphs of arbitrary high
clique-width (the $(n \times n)$-grids yield such a family of graphs,
see
\cite{MakowskyR99}).

\par \medskip \par \noindent
\textit{Proof of the claim:}
Let us show first how we can encode any finite set of elements
of~$\interpretation_R(t_R)$ by a pair of finite subsets of~$\dom(t_R)$
in such a way that the membership relation is ``definable''.

Let~$E$ be a finite set of vertices of~$\interpretation_R(t_R)$.  Each
vertex of~$\interpretation_R(t_R)$ is a finite set of nodes of~$t_R$
and therefore it makes sense to define $D_E$ as the union of all the
elements in~$E$.  Furthermore, let us chose~$I_E$ to be an element
of~$\interpretation_R(t_R)$ which is connected to all elements of~$E$
and to none of the elements of~$\mathcal{P}(D_E)\setminus E$ (such an element
exists since~$\interpretation_R(t_R)$ is the random graph). From $D_E$
and~$I_E$ one can easily reconstruct the set~$E$. More precisely,
let~$X$ be an element of~$\interpretation_R(t_R)$. Then $X$ belongs
to~$E$ if and only if $t_R$ models~$\delta(X)\wedge X\subseteq
D_E\wedge \Psi(X,I_E)$.

Let now~$G$ be a non-oriented finite graph.
Since~$\interpretation_R(t_R)$ is the random graph, the graph~$G$
appears as an induced subgraph of~$\interpretation_R(t_R)$
(cf. \cite{Hodges93}).  Let~$V$ be the set of vertices
of~$\interpretation_R(t_R)$ inducing this subgraph.

For each subset~$F$ of~$V$, one can construct an element~$v_F$
of~$\interpretation_R(t_R)$ which is connected to all the vertices
in~$F$ and to no vertex in~$V\setminus F$.  Knowing~$V$, this element~$v_F$
completely characterizes~$F$.  Let now~$V'$ be the set of all
the~$v_F$'s for all subsets $F$ of~$V$.

Let~$t_G$ be the tree~$t_R$ extended with markings
describing~$D_V,I_V,D_{V'}$ and~$I_{V'}$.  Using the trick mentioned
above, we can define the formula~$X\in V$ (similarly $X\in V'$) to
be~$\delta(X)\wedge X\subseteq D_V\wedge \Psi(X,I_V)$.  We now want to
finite sets interpret $\generatorF(G)$ in~$t_G$.  Obviously, we can
identify the elements of~$\generatorF(G)$ with the elements of~$V'$.
Pursuing this idea, we define the
interpretation~$\interpretation'=(\delta'(X),\Psi'(X,Y),\Phi_\subseteq(X,Y))$
in the following way. The universe is defined by~$\delta'(X)=X\in V'$.
The subset relation is defined by:
\begin{align*}
	\Phi_\subseteq(X,Y)&=\forall Z\in
V. \Psi(Z,X)\rightarrow\Psi(Z,Y)
\end{align*}
Finally the edge relation is
defined by:
\begin{align*}
\Psi'(X,Y)&= \mathit{Singleton}(X)\wedge\mathit{Singleton}(Y)\\
	&\wedge\exists X',Y'\in V. \Psi(X',X)\wedge\Psi(Y',Y)\wedge(\Psi(X',Y'))
\end{align*}
where~$\mathit{Singleton}(Z)$ stands for~$Z \in V'\wedge \exists!Y\in
V.\Psi(Y,Z)$ and $\exists!$ abbreviates ``there exists one and only
one''.

Using the properties linking~$V$ and~$V'$, it is not difficult to see
that~$\interpretation'(t_G)$ is (up to
isomorphism)~$\generatorF(G)$. Furthermore, $\interpretation'$ does
not depend on~$G$.  This finishes the proof of the claim and hence the
proof of the theorem.
\end{proof}

\subsection{Intrinsic definability}
\label{subsection:intrinsic-regularity}

Our last application of Theorem~\ref{tHeorem:main-result} concerns
``intrinsic definability'' of relations. This notion is the natural
adaption of the notion of intrinsic regularity for automatic
structures \cite{KhoussainovRS04}. An automatic structure may have (up
to isomorphism) several different presentations. These presentations
can have different properties in the following sense: It might be
possible to add a relation to the structure that is regular in one
presentation but is not regular in the other presentation.

Consider, for example, the structure $\nn$, i.e., the natural numbers
with the successor relation. One automatic presentation is to use
binary encodings of the numbers but it is also possible to use unary
encoding. If we now add the predicate ``being a power of $2$'', i.e.,
the set $\{2^n \mid n \in \nats\}$, then this predicate is certainly
not regular for the unary encoding but it is in the binary encoding
(it corresponds to the set of all words of the form $10^*$). This
means that this predicate is not intrinsically regular for $\nn$
because it is regular in some presentation but not in all.

Accordingly, a relation is called intrinsically regular for a
structure if it is regular in all automatic presentation of the
structure. In \cite{KhoussainovRS04} this notion is studied and the
question of a logical characterization of intrinsically regular
relations is raised.

In \cite{Barany06} it is shown that for the structure $(\{0,1\}^*,
S_0, S_1, \sqsubseteq, \mathit{el})$ (recall
Example~\ref{ex:MSO-to-FO} above) each relation is either intrinsically
regular or intrinsically non-regular, i.e., either it is regular in
every presentation or non-regular in every presentation of the
structure. Such a result can be used as a tool to show that certain
structures are not automatic, which is a difficult task
(cf. \cite{BlGr00}). If we add a relation to the above structure and
show that it is not regular in the natural automatic presentation,
then we know that the structure extended by this relation has no
automatic presentation at all.

In this subsection we show a stronger result for another structure. In
terms of automatic structures we show that for $\generatorF(\nn)$ each
relation is intrinsically regular or intrinsically non-regular for
every \emph{tree-automatic} presentation of $\generatorF(\nn)$.

For this, we adapt the notions to our setting. That is, intrinsic
definability considers relations that are definable in every possible
presentation of a structure by a finite sets interpretation from a
fixed tree $t$. If we, for example, fix this tree to be $\nn$,
then this corresponds to intrinsic regularity for automatic
structures.

Note that, in contrast to the previous sections, we
now explicitly consider the presentations of elements of a structure,
i.e., we distinguish different codings of the same structure.

\newcommand{\domstructure}{\mathcal{T}}

\begin{definition}
Given a structure~$\domstructure$, a
$\domstructure$-\intro{presentation} of a structure $\structure$ of
universe $\universe$ is an injection~$f$ from $\universe$ to the
finite subsets of~$\domstructure$ such that the set~$f(\universe)$ as
well as the image by~$f$ of each relation~$R$ of~$\structure$ are
WMSO-definable on $\domstructure$.  That is, there is a formula
$\phi^f_{\universe}(X)$ over $\domstructure$ defining the image of
$\universe$ under $f$, and for each relation $R$ of $\structure$ there
is a formula $\phi^f_{R}(X_1, \ldots,X_{r})$ using the signature of
$\domstructure$ such that for all $u_1, \ldots, u_{r} \in \universe$,
\[
(u_1, \ldots, u_{r}) \in R\quad\mbox{ iff }\quad \domstructure \models
\phi_R^f(f(u_1), \ldots, f(u_{r})),
\]
where $r$ denotes the arity of $R$.
\end{definition}
Given such a $\domstructure$-presentation $f$ of $\structure$, it
might be possible to add relations to $\structure$ such that $f$ is
still a $\domstructure$-presentation of this extended
structure. Such relations are called \intro{definable in $f$}, i.e.,
$R'$ is called definable in $f$ if $f$ is a
$\domstructure$-presentation of $\structure$ extended by the
relation $R'$.

Note that to a $\domstructure$-presentation $f$ of a structure
$\structure$ we can directly associate a finite sets interpretation
$\interpretation_f$ sending~$\domstructure$
to $\structure$ up to isomorphism (the isomorphism being~$f$).
An additional relation $R'$ is definable in $f$ if we can add to
$\interpretation_f$ a formula defining $R'$.

If $\structure$ is the weak powerset structure
$\generatorF(\domstructure)$ of $\domstructure$, then there is a
canonical $\domstructure$-presentation given by the identity
mapping. We refer to this $\domstructure$-presentation as the
\intro{standard presentation} of $\generatorF(\domstructure)$.

The following lemma states that the ``intrinsically definable''
relations of $\generatorF(t)$ for a tree $t$ are exactly those that
are regular in the standard presentation of $\generatorF(t)$.

%%%%%%%%%%%%%%%%%%%%%%%%%%%%%% LEMMA %%%%%%%%%%%%%%%%%%%%%%%%%%%%%%
\begin{lemma} \label{lemma:standard-regular}
Let $t$ be a tree and $R$ be a relation over $\generatorF(t)$. Then $R$
is definable in the standard presentation of $\generatorF(t)$ iff
$R$ is definable in all $t'$-presentations $f$ of $\generatorF(t)$ for
all trees $t'$.
\end{lemma}
%%%%%%%%%%%%%%%%%%%%%%%%%%%%%% LEMMA %%%%%%%%%%%%%%%%%%%%%%%%%%%%%%
\begin{proof}
Obviously, if $R$ is definable in all $t'$-presentations $f$ of
$\generatorF(t)$ for all trees $t'$, then $R$ is in particular
definable in the standard presentation.

For the other direction, let $R$ be a relation of arity $r$ that is
definable in the standard presentation of $\generatorF(t)$ and let
$\Phi_R(X_1, \ldots, X_r)$ be the defining formula, i.e., $\Phi_R$ is
a formula over the signature of $t$ such that $t \models \Phi_R(U_1,
\ldots, U_r)$ iff $(U_1,
\ldots, U_r) \in R$ for all $U_1, \ldots, U_r \subseteq \dom(t)$.
According to Proposition~\ref{proposition:powerset-generator} we can construct
an FO-interpretation $\interpretation_1$ such
that~$\interpretation_1(\generatorF(t))$ is the
structure~$\generatorF(t)$ augmented with the relation~$R$.

Let now $f$ be a $t'$-presentation of $\generatorF(t)$ and let
$\interpretation_f$ be the finite sets interpretation with
$\interpretation_f(t') =
\generatorF(t)$. Then~$(\interpretation_1\circ\interpretation_f)(t') =
\interpretation_1(\generatorF(t))$, witnessing the 
definability of~$R$ in the~$t'$-presentation~$f$ of~$\generatorF(t)$.
\end{proof}
The dual of Lemma~\ref{lemma:standard-regular}, where definable is
replaced by not definable, is not true in general.  This is already
the case for instance for $t = t' = \cbt$, i.e., there is a relation
$R$ which is not definable in the standard presentation of
$\generatorF(\cbt)$ but is definable in some $\cbt$-presentation of
$\generatorF(\cbt)$.

Consider, for example, the relation $R(x,y)$ that holds if $x$ is on
the leftmost branch, $y$ is on the rightmost branch, and $x$ and $y$
are on the same level. It is not difficult to see that this relation
is not WMSO-definable in $\cbt$. Hence, if we transfer $R$ to
singletons on $\generatorF(\cbt)$, it is not definable in the standard
presentation of $\generatorF(\cbt)$.

On the other hand, one can find a WMSO-interpretation
$\interpretation_2$ with $\interpretation_2(\cbt) \sim (\cbt,R)$. The
finite sets interpretation $\generatorF \circ \interpretation_2$
defines a $\cbt$-presentation of $\generatorF(\cbt)$ in which $R$
(transferred to singletons) is regular. The construction of
$\interpretation_2$ is not very difficult. It suffices to redefine
$\cbt$ inside itself such that the corresponding vertices of the
leftmost and the rightmost branch are located close to each
other, as for example done by the following mapping (where $w$ is any
non-empty word over $\{0,1\}$):
\[
0^n \mapsto 0^n, \;\; 1^n \mapsto 0^n1, \;\; 0^n1w \mapsto 0^n11w, \;\;
1^n0w \mapsto 0^n10w.
\]
The reader can verify that the two successor relations and the
relation $R$ are WMSO-definable on this coding of $\cbt$.

However, in the particular case of~$t=\nn$ and~$t'=\cbt$ such
a converse of Lemma~\ref{lemma:standard-regular} does hold as expressed
in the following theorem. 
%%%%%%%%%%%%%%%%%%%%%%%%%%%%%% THEOREM %%%%%%%%%%%%%%%%%%%%%%%%%%%%%%
\begin{theorem}\label{tHeorem:intrisic}
If $R$ is definable in some $\cbt$-presentation $f$
of~$\generatorF(\nn)$, it is definable in every
$\cbt$-presentation of~$\generatorF(\nn)$.
\end{theorem}
%%%%%%%%%%%%%%%%%%%%%%%%%%%%%% THEOREM %%%%%%%%%%%%%%%%%%%%%%%%%%%%%%
As we can expect, by application of Theorem~\ref{tHeorem:main-result}, we
will obtain a WMSO-in\-ter\-pre\-tation sending~$\cbt$
to~$\nn$. The two following lemmas study this kind of
interpretations and how they preserve definability.
%%%%%%%%%%%%%%%%%%%%%%%%%%%%%% LEMMA %%%%%%%%%%%%%%%%%%%%%%%%%%%%%%
\begin{lemma}\label{lemma:d2d1}
Let~$\interpretation_2$ be a WMSO-interpretation sending $\cbt$
to~$\nn$ and $f_2$ be the injection from~$\nats$ to~$\{0,1\}^*$
witnessing the isomorphism.  Then there exists naturals~$m,n>0$ and
words~$u,v,w_0,\dots,w_{n-1} \in \{0,1\}^*$ such that for all
naturals~$k \ge 0$ and~$p\in\{0,\dots,n-1\}$,
\begin{align*}
f_2(m+kn+p)&=uv^{k}w_p\ .
\end{align*}
\end{lemma}
%%%%%%%%%%%%%%%%%%%%%%%%%%%%%% LEMMA %%%%%%%%%%%%%%%%%%%%%%%%%%%%%%
\begin{proof}
The general idea behind the proof is that the elements from the image
of $f_2$ cannot be spread arbitrarily in $\cbt$ because the
successor relation from $\nn$ has to be definable in WMSO and hence
must be recognizable by an automaton.

We denote by $U \subseteq \{0,1\}^*$ the image of $f_2$ and by $V$ the
closure of $U$ by prefix. The set $V$ defines a subtree of $\cbt$. We
now augment $V$ by markings containing information on the successor
relation in such a way that these markings are definable in WMSO.
Hence, the marked tree is regular, i.e., it has only finitely
many non-isomorphic subtrees (see \cite{Thomas97} for more information
on regular trees).  From this regular tree we can define the words
$u,v,w_0,\dots,w_{n-1}$.

Let $\Phi_{\suc}(x,y)$ be the formula of $\interpretation$ that defines the
successor relation $\suc$ of $\nn$, and let $\auto_{\suc}$ be the
equivalent tree automaton.

As $\suc$ is deterministic, one can easily show that $V$ contains only
one infinite branch $B$. Otherwise, the relation $\suc$ has to jump
infinitely often between two infinite branches leading to a
contradiction as $\auto_{\suc}$ would also accept pairs of nodes that
are not in $\suc$. The argument is the same as in the proof of
Proposition~\ref{proposition:free-monoid} where it is shown that
$(\nats,+)$ is not WMSO-interpretable in any tree $t$.  Furthermore,
this branch $B$ is WMSO-definable.

Consider two nodes $x,y \in U$ such that $\Phi_{\suc}(x,y)$ is
satisfied, i.e., $f_2^{-1}(y)$ is the successor of $f_2^{-1}(x)$. We
can describe how to get from $x$ to $y$ by a pair of words
$(z_x,z'_x)$ over $\{0,1\}$ meaning that $x = x'z_x$ and $y = x'z'_x$
for the greatest common ancestor $x'$ of $x$ and $y$. Again,
using the determinism of $\suc$ one can show that the length of these
words $z_x,z'_x$ is bounded by some constant derived from the size of
$\auto_{\suc}$. Hence, we can mark the vertices from $U$ by this
information (using sets $X_{z,z'}$ with $x \in X_{z,z'}$ iff
$(z_x,z_x') = (z,z')$). Obviously, this marking is WMSO-definable.

The last information that we attach to $V$ is for each node $x \in B$
the word $b_x \in \{0,1\}^*$ such that the node $xb_x$ is the smallest
node in $U$ (smallest referring to the position in $\nn$) such
that all nodes bigger than $xb_x$ are below $x$, i.e., for all $y \in
U$, if $f_2^{-1}(xb_x) < f_2^{-1}(y)$, then $x \sqsubseteq y$. The length
of these $b_x$ is bounded because the relation that associates to each
$x$ the node $b_x$ is WMSO-definable and deterministic. Hence, the
marking of the nodes in $B$ by using sets $X_b$ with $x \in X_b$ iff
$b_x = b$ is WMSO-definable.

The resulting tree $t$, consisting of the nodes in $V$ with the markings
described above is WMSO-definable and hence regular. Let $u,v \in
\{0,1\}^*$ such that $u, uv \in B$ and the subtrees of $t$ rooted at $u$
and $uv$ are isomorphic. Let $m = f_2^{-1}(ub_u)$, $m' =
f_2^{-1}(uvb_{uv})$, and define $n = m' - m$. For $p \in \{0, \ldots,
n-1\}$ let $w_p \in \{0,1\}^*$ be such that $f_2(m + p) = uw_p$. By the
choice of $m$, such a $w_p$ always exists. In particular, $w_0 = b_u$.

By the choice of $v$, we know that $f_2(m +kn) = u v^{k} w_0$ for all
$k \ge 0$. Furthermore, as $t$ is marked by the information on how to
get from one node in $U$ to its successor, we know that the ways to
get from $f_2(m + kn + p)$ to $f_2(m+ kn + p + 1)$ are the same for all
$k$. Hence, $f_2(m + kn + p) = uv^{k}w_p$.
\end{proof}

\begin{lemma} \label{lemma:d2d1-definable}
Let~$\interpretation_2$ be a WMSO-interpretation sending $\cbt$
to~$\nn$ and $f_2$ be the injection from~$\nats$ to~$\{0,1\}^*$
witnessing the isomorphism. If $R$ is a relation over
finite subsets of~$f_2(\nats)$ WMSO-definable in~$\cbt$,
then its inverse image under~$f_2$ is
WMSO-definable in~$\nn$.
\end{lemma}
%%%%%%%%%%%%%%%%%%%%%%%%%%%%%% LEMMA %%%%%%%%%%%%%%%%%%%%%%%%%%%%%%
\begin{proof}
The formula $\Psi_R(X_1, \ldots, X_r)$ defining~$R$ can be represented
by a tree automaton $\auto_R$ with state set $Q_R$. Using
Lemma~\ref{lemma:d2d1}, this automaton can be simulated by an
automaton $\auto'_R$ on $\nn$ as we show in the following. On $\nn$
automata and WMSO have the same expressive power and hence the
construction of $\auto'_R$ suffices to prove the lemma.

Let $u,v,w_0, \ldots, w_{n-1}$ be as in Lemma~\ref{lemma:d2d1}.  The
states of $\auto_R'$ correspond to partial runs of $\auto_R$ on a
finite subtree of $\cbt$ that
\begin{itemize}
\item is rooted at $\epsilon$ and induced by the elements $f_2(0),
  \ldots, f_2(m-1)$ for the first segment of $\nn$,
\item rooted at $uv^{k}$ and induced by the words $v, w_0, \ldots,
  w_{n-1}$ for the following segments of $\nn$.
\end{itemize}
The corresponding parts of $\cbt$ on which $\auto_R'$ has to simulate
a run of $\auto_R$ are depicted in
Figure~\ref{figure:regular-subtree} (for $n=2$ and $m=3$).

For the formal definitions, let $U_0$ be the set of all nodes that are
prefix of $u$ or of some $f_2(i)$ for $0 \le i < m$, and let $U_1$ be
the set of all nodes that are prefix of $v$ or one of $w_0, \ldots,
w_{n-1}$. The set $U_0$ corresponds to the upper finite tree in
Figure~\ref{figure:regular-subtree} surrounded by a dashed line, and
the set $U_1$ to the finite tree rooted at $u$.

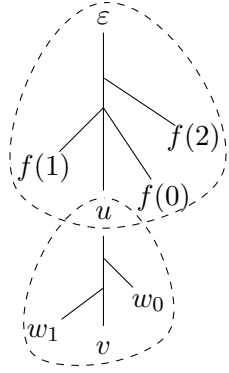
\begin{figure}
\begin{picture}(40,50)(60,-60)
\gasset{Nw=6,Nh=6,AHnb=0,Nframe=n}
\node[Nw=4,Nh=4](n0)(84.0,-12.0){$\varepsilon$}
\node(u)(84.0,-38.0){$u$}
\node(v)(84.0,-56.0){$v$}
\drawedge(n0,u){}
\drawedge(u,v){}

\node[Nw=0,Nh=0](n3)(84.0,-24.0){}
\node(n4)(76.0,-32.0){$f(1)$}
\node[Nw=0,Nh=0](n5)(84.0,-20.0){}
\node(n6)(96.0,-28.0){$f(2)$}
\node(n7)(92.0,-36.0){$f(0)$}
\drawedge(n5,n6){}
\drawedge(n3,n4){}
\drawedge(n3,n7){}
\drawccurve[Nframe=y,dash={1 1}](84,-10)(72,-34)(84,-40)(100,-32)

\node[Nw=0,Nh=0](n8)(84.0,-48.0){}
\node(n9)(76.0,-54.0){$w_1$}
\drawedge(n8,n9){}
\node[Nw=0,Nh=0](n11)(84.0,-44.0){}
\node(n10)(90.0,-50.0){$w_0$}
\drawedge(n11,n10){}
\drawccurve[Nframe=y,dash={1 1}](84,-36)(74,-56)(84,-58)(93,-53)
\end{picture}
\caption{Regular subtree of $\cbt$ induced by the domain of the
interpretation $\interpretation_2$ from
Lemmas~\ref{lemma:d2d1-definable} and \ref{lemma:d2d1}. The part
from $u$ to $v$ is iterated.} \label{figure:regular-subtree}
\end{figure}

The automaton $\auto'_R$ reads a word $\alpha \in (\{0,1\}^r)^\omega$
and has to decide if this labeling transferred by $f_2$ to $\cbt$
corresponds to a tuple of sets in $R$. To simulate a run of $\auto_R$
it guesses partial runs, starting with a partial run on $U_0$, and
then continuing with $U_1$, the periodic part of the tree.

More formally, it starts by guessing a pair $(\rho_0,\lambda_0)$ with
mappings $\rho_0: U_0 \rightarrow Q_R$ and $\lambda_0: \{f_2(0),
\ldots, f_2(m-1)\}
\rightarrow \{0,1\}^r$ such that $\rho_0$ corresponds to a partial run
of $\auto_R$ on $U_0$ with labels corresponding to $\lambda_0$. In the
next steps, $\auto'_R$ verifies if the guessed labeling is correct,
i.e., if $\alpha(i) = \lambda_0(f_2(i))$. When reaching position $m$,
$\auto'_R$ guesses a new pair $(\rho_1,
\lambda_1)$, now with mappings $\rho_1: U_1 \rightarrow Q_R$ and
$\lambda_1: \{w_0, \ldots, w_{n-1}\} \rightarrow \{0,1\}^r$ such that
$\rho_1$ is a possible continuation of $\rho_0$ on the subtree rooted at
$u$ that is shown in Figure~\ref{figure:regular-subtree} and labeled
according to $\lambda_1$. The guessed labeling is again verified on
the segment $m, \ldots, m + n-1$ and then a pair $(\rho_2, \lambda_2)$
of the same type as $(\rho_1,\lambda_1)$ is guessed, and so on. The
automaton accepts if the concatenation of the guessed partial runs on
the path $uv^\omega$ satisfies the acceptance condition of
$\auto_R$. For this to work, the guessed partial runs have to be such
that they can be continued to accepting runs on the ``blank parts'' of
$\cbt$, i.e., those infinite subtrees that do not contain a node
from the image of $f_2$.
\end{proof}
Using this Lemma we can prove Theorem~\ref{tHeorem:intrisic}.

\medskip
\begin{proof}[Proof of Theorem~\ref{tHeorem:intrisic}]
Assuming a $\cbt$-presentation~$f$ of~$\generatorF(\nn)$
and~$\interpretation$ the corresponding finite sets interpretation,
one obtains by Corollary~\ref{corollary:main-general} a
WMSO-interpretation $\interpretation_2$ with $\interpretation_2(\cbt)
= \nn$.  Let $f_2: \nats \rightarrow \{0,1\}^*$ be the injection
witnessing this isomorphism and let~$\tilde{f_2}$ be its extension to
sets.

We now have two ways to obtain isomorphic copies of
$\generatorF(\Delta_1)$ from $\Delta_2$: by applying $\interpretation$
and by applying $\interpretation_2$ followed by $\generatorF$. These
two ways yield isomorphic structures and hence there is an isomorphism
$h$ sending $\interpretation(\Delta_2)$ to
$\generatorF(\interpretation_2(\Delta_2))$.  We obtain the following
picture, where dashed arrows represent interpretations, while normal
arrows are for isomorphisms:
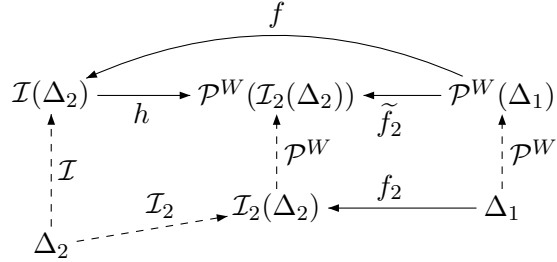
\begin{center}
\begin{picture}(70,39)
\gasset{Nframe=n,Nadjust=w,Nh=5}
\node(ID2)(5,25){$\interpretation(\Delta_2)$}
\node(GI2D2)(35,25){$\generatorF(\interpretation_2(\Delta_2))$}
\node(GD1)(65,25){$\generatorF(\Delta_1)$}
\node(D2)(5,5){$\Delta_2$}
\node(I2D2)(35,10){$\interpretation_2(\Delta_2)$}
\node(D1)(65,10){$\Delta_1$}

\drawedge[ELside=r,ELpos=40](ID2,GI2D2){$h$}
\drawedge(GD1,GI2D2){$\widetilde{f_2}$}
\drawedge[curvedepth=-8,ELside=r](GD1,ID2){$f$}
\drawedge[ELside=r](D1,I2D2){$f_2$}

\drawedge[dash={1 1}{0.0},ELside=r](D2,ID2){$\interpretation$}
\drawedge[dash={1 1}{0.0},ELside=r](I2D2,GI2D2){$\generatorF$}
\drawedge[dash={1 1}{0.0},ELside=r](D1,GD1){$\generatorF$}
\drawedge[dash={1 1}{0.0}](D2,I2D2){$\interpretation_2$}
\end{picture}
\end{center}
We show that we can define the isomorphism $h$ on $\Delta_2$ by a
WMSO-formula.  To understand this, consider a finite set $U$ of
naturals, i.e., an element of $\generatorF(\Delta_1)$. This set $U$
corresponds to two sets $X$ and $Y$ of nodes of $\Delta_2$, namely $X
= f(U)$ and $Y = \tilde{f_2}(U)$. The isomorphism $h$ relates these
two sets, i.e., $h(X) = Y$. This relation can be defined in WMSO using
the formula $\code$ that we obtain from
Theorem~\ref{tHeorem:main-result} and that is used to construct
$\interpretation_2$ in Corollary~\ref{corollary:main-general}.  The
formula $\code(X,x)$ relates each subset of $\cbt$ that is an atom in
$\generatorF(\nn) = \interpretation(\cbt)$ to a single node $x$ of
$\cbt$. Assume that the set $X$ represents the atom $\{n\}$. Then the
unique $x$ such that $\code(X,x)$ is satisfied represents $n$ in
$\interpretation_2(\cbt) = \nn$ (by the construction of
$\interpretation_2$ in the proof of
Corollary~\ref{corollary:main-general}), and the singleton $\{x\}$
represents the atom $\{n\}$ in
$\generatorF(\interpretation_2(\cbt))$. We get that $h(X) = x$, i.e.,
the formula $\code$ defines the isomorphism $h$ on the level of
atoms. It is easy to extend this to sets:
\begin{align*}
\phi_h(X,Y) &= \forall y (y \in Y \leftrightarrow \exists Z
(\Phi_{\subseteq}(Z,X) \land \code(Z,y))).
\end{align*}
Let now $R$ be a relation over~$\generatorF(\nn)$ which is definable
in $f$. This means that~$f(R)$ is WMSO-definable in~$\cbt$. Since~$h$
is WMSO-definable, it follows that~$h(f(R))$ is also WMSO-definable
in~$\cbt$.  Finally, by Lemma~\ref{lemma:d2d1-definable} we obtain
that $\tilde{f_2}^{-1}(h(f(R)))$ is WMSO-definable in~$\nn$. Since
$\tilde{f_2}^{-1}\circ h\circ f$ is obtained as a composition of
isomorphisms, it is an automorphism of~$\generatorF(\nn)$. Remark now
that the identity is the only automorphism of~$\generatorF(\nn)$ (a
property inherited from~$\nn$). It follows that
$\tilde{f_2}^{-1}(h(f(R)))$ equals~$R$. And consequently~$R$ is
definable in the standard presentation.  By
Lemma~\ref{lemma:standard-regular} it is definable in every
presentation.
\end{proof}

\section{Proof of the main result}
\label{section:result-proof}
The proof of Theorem~\ref{tHeorem:main-result} is rather complex and
split into several parts.  In
Subsection~\ref{subsection:result-intro}, we introduce the key notions
used afterwards while we make the scheme of the proof more
precise. This will also be the occasion for explaining the content of
Subsections~\ref{sub:important},
\ref{sub:standard}, \ref{sub:inf-branches},
\ref{sub:car-parking}.
In Subsection~\ref{sub:mainproof}, things are put together and the
proof of Theorem~\ref{tHeorem:main-result} is finally given.

%-------------------------------------------------------------------------------
\subsection{First definitions and presentation of the proof}\label{subsection:result-intro}
%-------------------------------------------------------------------------------

We assume from now that a finite sets interpretation $\interpretation
=(\delta(X),\phi_\preceq(X,Y))$ is fixed.  Along the whole proof
we use a tree~$t$ together with a set $E$ and the isomorphism~$f$
that are assumed to satisfy the equality
\begin{equation*}
f(\powerF(E))=\interpretation(t)~.
\end{equation*}
The reader must keep in mind that none of the constructions we perform
makes use of $t$, $E$, or $f$.  Hence, the result will hold for any
such tree, set, and isomorphism.  This lightens the presentation of
the proof by avoiding to systematically quantify over those objects.

We consider the set $\atoms$ of finite subsets of~$t$ representing
atoms of the powerset lattice, i.e.,
\begin{align*}
\atoms&=\set{f(\set{u})~:~u \in E}\ .
\end{align*}
The set $\atoms$ can be defined as the set of finite subsets
of~$t$ which are minimal --- for the~$\phi_\preceq$ formula seen as
an ordering --- and distinct from the minimal element itself (which
is~$f(\emptyset)$).  This description can be done in weak monadic
second-order logic.  Hence~$\atoms$ is regular in~$t$ and there
exists an automaton
\begin{align*}
\auto_\atoms=&(Q_\atoms,\qin_\atoms,
	\delta_\atoms,\Omega_\atoms)
\end{align*}
accepting the language~$\atoms$.  We also consider the binary
relation $\membership$ over~$\interpretation(t)$ defined as the image
under~$f$ of the~$\in$ relation in~$\powerF(E)$, i.e.,
\begin{align*}
\membership&=\set{(f(\set{u}),f(V))~:~u\in V \subseteq E \mbox{ and
    $V$ finite}}\ .
\end{align*}
This $\membership$ relation is also definable in weak monadic
second-order logic, and consequently is regular. We fix
\begin{align*}
\auto_\membership&=(Q_\membership,
	\qin_\membership,
	\delta_\membership,
	\Omega_\membership)
\end{align*}
to be an automaton recognizing the relation~$\membership$.

Recall that the theorem we want to prove claims the existence of a
formula $\code(X,x)$ such that the corresponding relation is an
injection from~$\atoms$ into $\dom(t)$.

Our goal in the construction of~$\code$ is to uniquely attach to
each~$X$ in $\atoms$ an element in~$\dom(t)$ in a WMSO-definable way.
As a first approximation, in Subsections~\ref{sub:important},
\ref{sub:standard}
and~\ref{sub:inf-branches}, we define a mapping $\funind$ which
assigns to each~$X$ in~$\atoms$ a node in~$\dom(t)$.  Though the
$\funind$ mapping is not in general an injection from~$\atoms$ into
$\dom(t)$, it does not either concentrate a lot of indices in the same
area of the tree~$t$.  Formally, if we set~$D(x)$ for~$x\in t$ to be
the cardinality of~$\funind^{-1}(x)$, then by
Lemmas~\ref{lemma:sparsity} and~\ref{lemma:branch-sparsity}, $D$
happens to be a sparse distribution (see
Definition~\ref{definition:sparse} below).
Subsection~\ref{sub:car-parking} is dedicated to the study of sparse
distributions. The central lemma of this part,
Lemma~\ref{lemma:car-parking}, establishes that, given elements
concentrated according to a sparse distribution, we can uniformly
redistribute them in~$\dom(t)$ in a unique WMSO-definable way. Applied
to our case, this means that the $\funind$ mapping can be transformed
into an injection by use of WMSO-formulas. And this last step is used
in Subsection~\ref{sub:mainproof} for terminating the proof of
Theorem~\ref{tHeorem:main-result}.

The key definition connecting the two main parts of the proof
(definition of the $\funind$ mapping and turning it into an injection)
is the notion of sparsity. This definition requires the notion of
zone.  A \intro{zone}~$Z$ in~$t$ is a connected --- where $t$ is seen
as an non-oriented graph --- subset of~$\dom(t)$.  That is, $Z$
contains a minimal element w.r.t. to the prefix ordering, and
whenever~$x\sqsubseteq y\sqsubseteq z$ for~$x,z$ in~$Z$, then also
$y\in Z$.  A zone~$Z$ is completely characterized by its least
element~$x$, and by the minimal elements~$x_1,\dots,x_n$ that are
below~$x$ and not in~$Z$. The elements~$\{x,x_1,\dots,x_n\}$ are
called the \intro{frontier} of the zone.  Given nodes
$x,x_1,\dots,x_n$ of~$t$ such that the~$x_i$ are pairwise incomparable
and $x\sqsubseteq x_i$ for all $i$, we define $\st{x,x_1,\dots,x_n}$
to be the set of nodes~$y$ such that~$x\sqsubseteq y$ and
$x_i\not\sqsubseteq y$ for all~$i \in [n]$ where $[n]$ denotes the set
$\{1, \ldots n\}$.  By construction, $\st{x,x_1,\dots,x_n}$ is the only
zone which has frontier~$\{x,x_1,\dots,x_n\}$.

\begin{definition}\label{definition:sparse}
A \intro{distribution} $D$ is a mapping from~$\dom(t)$
to~$\nats$.  For~$Z$ a finite zone, $D(Z)$ stands for~$\sum_{x\in
Z}D(x)$.  A distribution~$D$ is $K$-\intro{sparse} for
some~$K\in\nats$ if for every finite zone~$Z$ of frontier~$F$, $D(Z)\leq
\card Z+ K\card{F}$.
A distribution is \intro{strongly} $K$-\intro{sparse} if for any
finite zone~$Z$ of frontier~$F$, $D(Z)\leq K\card{F}$.
\end{definition}
Sparsity tells us that no finite zone contains more indices than its
size plus a factor linearly depending on the size of the frontier.

%-------------------------------------------------------------------------------
\subsection{Important nodes}\label{sub:important}
%-------------------------------------------------------------------------------
In order to construct the mapping $\funind$, given an element~$X$
of~$\atoms$, we first define the set $I(X)\subseteq\dom(t)$ of its
important nodes via combinatorial constraints.  Essentially, we try to
locate the places where ``important coding decisions'' are made by the
automaton~$\auto_\atoms$ when reading~$X$.  In the present Subsection, we
provide the key combinatorial lemmas concerning important nodes.

Then, depending on the shape of the set~$I(X)$ two cases are separated
and two distinct definitions of $\funind$ are given.  The first kind
of index is called standard index --- noted $\funsind(X)$ --- and is
the subject of Subsection~\ref{sub:standard}.  The other kind is called
branch index --- noted $\funbind(X)$ --- and is the subject of
Subsection~\ref{sub:inf-branches}.

Let us first introduce a convenient notation for studying
the behavior of the automata
$\auto_\atoms$ and~$\auto_\membership$ over zones:
For a zone $Z = \st{x,x_1, \ldots, x_n}$ and states $q,q_1,
\ldots, q_n \in Q_{\atoms}$, we denote by $\atoms(q,Z,q_1,
\ldots, q_n)$ the set of all $X \subseteq Z$ such that there exists $X'
\subseteq \dom(t)$ with
\begin{itemize}
\item $X = X' \cap Z$, and
\item $X'$ is accepted by $\auto_{\atoms}$ with a run $\rho$ such
  that $\rho(x) = q$ and $\rho(x_i) = q_i$ for all $i \in [n]$.
\end{itemize}
Similarly, for $q,q_1, \ldots, q_n \in Q_{\membership}$ we denote by
$\membership(q,Z,q_1, \ldots, q_n)$ the set of all pairs $(X,Y)$ with
$X,Y \subseteq Z$ such that there are $X',Y' \subseteq \dom(t)$ with
\begin{itemize}
\item $X = X' \cap Z$, $Y = Y' \cap Z$, and
\item $(X',Y')$ is accepted by $\auto_{\membership}$ with a run $\rho$
  such that $\rho(x) = q$ and $\rho(x_i) = q_i$ for all $i \in [n]$.
\end{itemize}

The definition of important nodes is then the following, where the
constant $\kindex$ is chosen to make the combinatorial arguments in
the subsequent lemmas work.
\begin{definition}
Let $\kindex = (2 \card{Q_{\membership}} +
1)\card{Q_{\atoms}}$. Given $X\in\atoms$, a node~$x\in
\dom(t)$ is called \intro{important} for~$X$ if
\begin{align*}
\card{\{Y\subseteq \st{x}~:~(X - \st{x}) \cup Y \in \atoms \}}> \kindex\ .
\end{align*}
We denote by~$I(X)$ the set of important nodes for~$X$.
\end{definition}
Hence a node~$x$ is important for~$X$ if there are many --- i.e., more
than $\kindex$ --- ways to modify~$X$ below~$x$ while remaining
in~$\atoms$. Intuitively, without knowing how $X$ looks like below
$x$, we cannot say much about which atom is coded because there are
too many possibilities left.  Remark that the set~$I(X)$ is by
definition prefix closed.  The fundamental property that we show in
Lemma~\ref{lemma:killing-lemma} is that for an important node $x$ of
$X$, the part of $X$ that is not below $x$ comes from a set of small
size. To prove this lemma we need its combinatorial core stated in the
following lemma.
%%%%%%%%%%%%%%%%%%%%%%%%%%%%%% LEMMA %%%%%%%%%%%%%%%%%%%%%%%%%%%%%%
\begin{lemma} \label{lemma:combinatorial} Let $\kcomb = 2 \card{Q_{\membership}} +
1$. For any two disjoint zones~$Z$ and~$Z'$ of respective frontiers~$F
= \{x, x_1, \ldots, x_n\}$ and~$F' = \{x', x_1', \ldots,
x_m'\}$, and all accepting runs~$\rho$ of~$\auto_{\atoms}$
\begin{align*}
\text{either} \quad &
 \card{\atoms(\rho(x), Z,\rho(x_1), \ldots, \rho(x_n))} < \kcomb\ ,
\\
\text{or} \quad &
 \card{\atoms(\rho(x'), Z',\rho(x_1'), \ldots, \rho(x_m'))} < \kcomb\ .
\end{align*}
\end{lemma}
%%%%%%%%%%%%%%%%%%%%%%%%%%%%%% LEMMA %%%%%%%%%%%%%%%%%%%%%%%%%%%%%%
\begin{proof}
It is sufficient for us to prove the result for two complementary
zones.  This comes from the fact that increasing a zone also increases
the number of possible projections w.r.t.\ a fixed run, i.e.,
$\card{\atoms(\rho(x), Z,\rho(x_1), \ldots, \rho(x_n))} \leq
\card{\atoms(\rho(y), Z'', \rho(y_1), \ldots, \rho(y_{\ell}))}$
for a zone $Z''$ of frontier $\{y, y_1, \ldots, y_{\ell}\}$ with $Z
\subseteq Z''$.  Hence, we will assume~$Z$ to be~$\st{\epsilon,x}$ and
$Z'$ to be~$\st{x}$ for some node~$x$.

Assume that for some $K \ge \kcomb$ we have distinct sets
$X_1,\dots,X_{K}$ in $\atoms(\rho(\epsilon), Z,\rho(x))$ and
distinct sets $X'_1,\dots,X'_{K}$ in $\atoms(\rho(x), Z')$.  Then,
for every~$i,j\in[K]$, let $Y_{i,j}$ be~$X_i\cup X'_j$.  As the union
of $Z$ and $Z'$ gives the whole domain of $t$, we have $Y_{i,j}\in
\atoms$ for all $i,j\in[K]$.

Let us now consider the set~$\mathit{Comb}$ of possible combinations
of the $Y_{i,j}$, combination in the sense of the relation~$\membership$.
More precisely, $A\subseteq \dom(t)$ is in $\mathit{Comb}$
if whenever $(Y,A) \in \membership$
holds for some atom~$Y$,
then $Y = Y_{i,j}$ for some $i,j$.
The cardinality
of~$\mathit{Comb}$ is~$2^{K^2}$. We now show by a combinatorial
argument that $\auto_{\membership}$ cannot distinguish all the
elements from $\mathit{Comb}$ because the amount of information that
can be passed between the two zones $Z$ and $Z'$ is limited by the
number of states in $Q_{\membership}$.

For this purpose, we define for each~$A\in\mathit{Comb},
$~$f_A:[K]\times Q_\membership\rightarrow\{0,1\}$
and~$g_A:Q_\membership\times[K]\rightarrow\{0,1\}$ by

\begin{align*}
f_A(i,q)&=\begin{cases}
	1&\text{if}~(X_i,A\cap Z)
		\in \membership(\qin_\membership, Z, q), \\
	0&\text{else}, \end{cases}\\
g_A(q,j)&=\begin{cases}
	1&\text{if}~(X'_j,A\cap Z')
		\in \membership(q, Z'), \\
	0&\text{else}.\end{cases}
\end{align*}
It is obvious that if two sets~$A,B\in\mathit{Comb}$ are such
that~$f_A=f_B$ and~$g_A=g_B$, then $(Y_{i,j},A) \in \membership$ iff
$(Y_{i,j},B) \in \membership$ for all $i,j \in [K]$.  This means,
by definition of~$\mathit{Comb}$, that~$A=B$.

However, there are only~$2^{2\card{Q_\membership}K}$ different
possible values for the pair~$f_A,g_A$. Hence
we obtain~$\card{\mathit{Comb}}\leq 2^{2\card{Q_\membership}K}$.
This contradicts~$\card{\mathit{Comb}}=2^{K^2}$.
\end{proof}

The following lemma shows that the possibilities to code an atom
`above' an important node are bounded.
%%%%%%%%%%%%%%%%%%%%%%%%%%%%%% LEMMA %%%%%%%%%%%%%%%%%%%%%%%%%%%%%%
\begin{lemma} \label{lemma:killing-lemma}
For each node $x$ we have
%\[
$
\card{\{X \cap \st{\epsilon,x}~:~X \in \atoms \mbox{ and } x \in
  I(X)\}} < \kindex.
$
%\]
\end{lemma}
%%%%%%%%%%%%%%%%%%%%%%%%%%%%%% LEMMA %%%%%%%%%%%%%%%%%%%%%%%%%%%%%%
\begin{proof}
We are aiming at a contradiction to Lemma~\ref{lemma:combinatorial} for
$Z = \st{\epsilon,x}$ and $Z' = \st{x}$.

For each $X$ with $x \in I(X)$ there are more than $\kindex$ many $Y
\subseteq \st{x}$ such that $X_{Y,x} := (X - \st{x}) \cup Y$ is in
$\atoms$. Since $\kindex = |Q_{\atoms}| \cdot \kcomb$ (with
$\kcomb$ from Lemma~\ref{lemma:combinatorial}), we can choose a state
$q_{X,x} \in Q_{\atoms}$ such that more than $\kcomb$ of these
$X_{Y,x}$ are accepted by $\auto_{\atoms}$ with a run that labels
$x$ with $q_{X,x}$. This means that $|\atoms(q_{X,x},\st{x})| \ge
\kcomb$.

Now, assume that there are $\kindex$ different $X \in \atoms$ with
$x \in I(X)$ that differ on $\st{\epsilon,x}$. Then there are at least
$\kcomb$ such sets $X_1, \ldots, X_{\kcomb}$ with $q_{X_1,x} = \cdots
= q_{X_{\kcomb},x} =: q$. In particular, we obtain
$|\atoms(\qin_{\atoms},\st{x},q)| \ge \kcomb$. Together with
$|\atoms(q,\st{x})| \ge \kcomb$ from above we obtain the desired
contradiction.
\end{proof}

%-------------------------------------------------------------------------------
\subsection{Standard indices} \label{sub:standard}
%-------------------------------------------------------------------------------
We now address the problem of computing~$\funind(X)$
for some atom~$X$
under the assumption that~$I(X)$ is \emph{not an infinite branch}
(the case when~$I(X)$ is an infinite branch is treated in
Subsection~\ref{sub:inf-branches}).
Since we call this case the standard case, we will
denote the index defined for such atoms $X$ by $\funsind(X)$.  The
simplest case is that $I(X)$ is totally ordered by~$\sqsubseteq$,
i.e., $I(X)$ is a finite path starting from the root.
We simply define~$\funsind(X)$ to be the last node on this path.
The other case corresponds to~$I(X)$ not being a finite path
nor an infinite branch. This corresponds to $I(X)$
not being totally ordered by~$\sqsubseteq$.
In this situation, we define~$\funsind(X)$ to be the first
node at which $I(X)$ splits into two paths.
Those two cases are unified in the following definition.
\begin{definition}
For $X \in \atoms$ such that $I(X)$ is not an infinite branch, the
\intro{index} of~$X$, written $\funsind(X)$, is the maximal element
in~$I(X)$ which is comparable to every element in~$I(X)$.
\end{definition}

As already mentioned, the intention of this definition is
that~$\funsind(X)$ roughly locates in the tree where the main information
concerning the atom coded by $X$ lies.  This location is far from
being precise, and many elements of $\atoms$ may have the same
index. However, we will see that it is possible to obtain a good
understanding of the repartition of the standard indices. The following lemma
gives precise bounds on the quantity of indices that may occur in a
zone, i.e., it states that the distribution assigning to each node $x$
the number of $X$ such that $\funsind(X) = x$ is sparse.
%%%%%%%%%%%%%%%%%%%%%%%%%%%%%% LEMMA %%%%%%%%%%%%%%%%%%%%%%%%%%%%%%
\begin{lemma}[sparsity]\label{lemma:sparsity}
There is a constant $\ksparse$ such that
$\card{\funsind^{-1}(Z)}\leq \card{Z}+ \ksparse \card{F}$
for every finite zone~$Z$ of frontier~$F$.
\end{lemma}
%%%%%%%%%%%%%%%%%%%%%%%%%%%%%% LEMMA %%%%%%%%%%%%%%%%%%%%%%%%%%%%%%
\begin{proof}
Denote the elements of the frontier of $Z$ by $x$ and $x_1, \ldots,
x_n$, i.e., $Z = \st{x,x_1, \ldots, x_n}$.  The proof of the lemma
consists of two steps. We first show that for atoms $X$ such that
$\funsind(X)$ is inside $Z$, the amount of information
located outside $Z$ is bounded.
More precisely, we first show for $M := \kindex \cdot |Q_{\atoms}|$
\begin{enumerate}[(a)]
\item $\card{\{X \cap \st{\epsilon,x}~:~ X \in \atoms \mbox{ and }
  \funsind(X)\in Z\}} < M$ and
\item $\card{\{X \cap \st{x_i}~:~ X \in \atoms \mbox{ and }
  \funsind(X)\in Z\}} < M$ for all $i \in
  [n]$.
\end{enumerate}
For (a) note that from $\funsind(X) \in Z$, the definition of
$\funsind$, and the prefix closure of $I(X)$ we obtain that $x \in
I(X)$. Therefore,
\begin{align*}
\{X \cap \st{\epsilon,x}~:~ X \in \atoms \mbox{ and }
\funsind(X)\in Z\} \\
\ \ \subseteq \{X \cap \st{\epsilon,x}~:~ X \in
\atoms \mbox{ and } x \in I(X)\}
\end{align*}
and (a) follows from Lemma~\ref{lemma:killing-lemma}.

For (b) we show that for each $X \in \funsind^{-1}(Z)$ and each $x_i$
there is a state $q \in Q_{\atoms}$ such that $X \cap \st{x_i} \in
\atoms(q,\st{x_i})$ and $\card{\atoms(q,\st{x_i})} <
\kindex$. From this, (b) follows because each $X \cap \st{x_i}$ comes
from one of at most $\card{Q_{\atoms}}$ many sets of size less
than $\kindex$.  We distinguish two cases.

If $x_i \notin I(X)$, then
we take $q$ to be the state at $x_i$ in an accepting run of
$\auto_{\atoms}$ on $X$. From the definition of $I(X)$ we
immediately obtain the desired property.

Else, if $x_i \in I(X)$, by definition of standard indices
there must some~$y\in I(X)$ incomparable to~$x_i$,
the index of~$X$ being the deepest common ancestor of~$x_i$ and~$y$.
From the definition of important nodes for $X$ and from $\kindex =
\card{Q_{\atoms}} \cdot \kcomb$ we obtain that there exists a set
$Y \subseteq \st{y}$ such that $(X - \st{y}) \cup Y$ is in
$\atoms$ and is accepted with a run of $\auto_{\atoms}$ that
labels $y$ by a state $q'$ such that $\card{\atoms(q',\st{y})} \ge
\kcomb$.  Let $q$ be the state assumed at node $x_i$ by this run.
From Lemma~\ref{lemma:combinatorial} applied to the zones $\st{x_i}$,
$\st{y}$, and to the aforementioned run, we can conclude that
$\card{\atoms(q,\st{x_i})} < \kcomb$. The desired property follows
from $\kcomb \le \kindex$. This finishes the proof of (b).

After these preliminary considerations, we come back to the claim of
the lemma.
We denote the elements from $\{X \cap \st{\epsilon,x}~:~ \funsind(X)
\in Z\}$ by $X^1, \ldots, X^{M}$ and the elements from $\{X \cap
\st{x_i}~:~ \funsind(X) \in Z\}$ by $X_i^1, \ldots, X_i^{M}$
(the same element can be represented more than once, what is important
is that all elements are represented).

Now, consider the set $\mathit{Comb}$ of all combinations of
atoms from $\funsind^{-1}(Z)$ (in the same sense as in the proof
of Lemma~\ref{lemma:combinatorial}). A combination $A \in \mathit{Comb}$
is entirely characterized by the following objects
\begin{itemize}
\item the set $A \cap Z$,
\item the mapping $f_{A,x}: [M] \times Q_{\membership} \rightarrow \{0,1\}$
with
\[
f_{A,x}(j,q) =
\begin{cases} 1 \qquad& \text{if}\quad
(X^j, A \cap \st{\epsilon,x}) \in
\membership(\qin_\membership, \st{\epsilon,x}, q),\\
0&\text{else,}
\end{cases}
\]
\item and the mapping
$f_{A,x_i}: Q_{\membership} \times [M] \rightarrow \{0,1\}$ with
\[
f_{A,x_i}(q,j) =
\begin{cases} 1 \qquad& \text{if}\quad
(X_i^j, A \cap \st{x_i}) \in \membership(q, \st{x_i}),\\
0&\text{else.}
\end{cases}
\]
\end{itemize}
Thus, $\card{\mathit{Comb}} \le 2^{\card{Z}} \cdot 2^{M
\card{Q_{\membership}}} \cdot \prod_{i=1}^n
2^{\card{Q_{\membership}}M}$. For $\ksparse =
\card{Q_{\membership}} M$ we obtain
\[
2^{\card{\funsind^{-1}(Z)}} = \card{\mathit{Comb}} \le 2^{\card{Z} +
  |F| \ksparse}
\]
and hence $\card{\funsind^{-1}(Z)} \le \card{Z} + \ksparse \card{F}$.
\end{proof}

%-------------------------------------------------------------------------------
\subsection{Treatment of infinite branches} \label{sub:inf-branches}
%-------------------------------------------------------------------------------
It is possible that for some $X \in \atoms$ the set $I(X)$ of
important nodes is an infinite branch. For these $X$ we also develop a
notion of index, called $\funbind(X)$, and show that the distribution
obtained in this way is strongly sparse. Since the sum of a $K$-sparse
distribution and of a strongly $K'$-sparse distribution is a
$K+K'$-sparse distribution, we can add the indices corresponding to
infinite branches to the other indices without affecting the sparsity
of the induced distribution.

In this subsection, we call the infinite branches that are equal to
$I(X)$ for some atom~$X$ \intro{important branches}.  We start with
the helpful observation that the number of elements of $\atoms$
corresponding to the same important branch is bounded.
%%%%%%%%%%%%%%%%%%%%%%%%%%%%%% LEMMA %%%%%%%%%%%%%%%%%%%%%%%%%%%%%%
\begin{lemma} \label{lemma:bounded-branchindex}
For every important branch~$B$, $\card{I^{-1}(B)} < \kindex$.
\end{lemma}
%%%%%%%%%%%%%%%%%%%%%%%%%%%%%% LEMMA %%%%%%%%%%%%%%%%%%%%%%%%%%%%%%
\begin{proof}
If there are $\kindex$ different sets in $I^{-1}(B)$, then we can pick
a node $x$ on $B$ such that all these sets differ on the zone
$\st{\epsilon,x}$. Since $x$ is important for all $X$ in $I^{-1}(B)$,
we obtain a contradiction to Lemma~\ref{lemma:killing-lemma}.
\end{proof}

Our goal is to associate to every important branch~$B$
a node $\funvind(B)$ on $B$ such that
\begin{enumerate}[(1)]
\item at most $\kindex$ branches are mapped to the same node by~$\funvind$, and
\item if some $\funvind(B')$ is above $\funvind(B)$, then
  $\funvind(B)$ is not in~$B'$.
\end{enumerate}
Those two properties are established in Lemma~\ref{lemma:vind-properties}.
Then Lemma~\ref{lemma:branch-sparsity} uses those two properties for
concluding that~$\funvind\circ I$ has a strongly sparse distribution.

Our main tool for constructing~$\funvind$ is to
produce a well-founded order for branches.
For this, we define $\funrind(B)$ for every important branch~$B$ by
\[
\funrind(B) = \min\{x \in B:\exists X \subseteq \st{\epsilon,x},~
I(X) = B\}.
\]
Since we consider finite sets interpretations, $\funrind(B)$ is
always defined.
Lemma~\ref{lemma:killing-lemma} applied to the node~$\funrind(B)$
directly leads to the following lemma.
%%%%%%%%%%%%%%%%%%%%%%%%%%%%%% LEMMA %%%%%%%%%%%%%%%%%%%%%%%%%%%%%%
\begin{lemma} \label{lemma:bounded-rind}
For all nodes $x$, $\card{\funrind^{-1}(x)} < \kindex$.
\end{lemma}
%%%%%%%%%%%%%%%%%%%%%%%%%%%%%% LEMMA %%%%%%%%%%%%%%%%%%%%%%%%%%%%%%
The well-foundedness argument announced above is then the following.
\begin{lemma}\label{lemma:prec-wf}
For every important branch~$B$, there are finitely many important
branches~$B'$ such that~$\funrind(B')\sqsubseteq\funrind(B)$.
\end{lemma}
\begin{proof}
One has that $\funrind(B')\sqsubseteq\funrind(B)$ iff
$B'$ belongs to $\cup_{x\sqsubseteq
\funrind(B)}\funrind^{-1}(x)$. This set is finite by
Lemma~\ref{lemma:bounded-rind}.
\end{proof}

Now, we can define for a branch~$B$ the index~$\funvind(B)$
as being the first node in~$B$ below~$\funrind(B)$
which is not lying on an important branch strictly
inferior with respect to comparing the~$\funrind$ values.
Formally
\begin{align*}
\funvind(B)&=\min\{x \in B~:~\funrind(B)\sqsubseteq x,~\forall
 B'.~\funrind(B')\sqsubset\funrind(B)\rightarrow x\notin B'\}.
\end{align*}
This definition is sound thanks to Lemma~\ref{lemma:prec-wf}.
Furthermore, $\funvind$ and~$\funrind$ can be related in
the following way.
\begin{lemma}\label{lemma:vind-rind}
$\funvind(B)\sqsubseteq\funvind(B')$ implies
$\funrind(B)\sqsubseteq\funrind(B')$.
\end{lemma}
\begin{proof}
Assume~$\funvind(B)\sqsubseteq\funvind(B')$.
Since~$\funvind(B')\in B'$ we obtain $\funvind(B)\in B'$.
As by definition $\funrind(B)\sqsubseteq\funvind(B)$,
we also have~$\funrind(B)\in B'$. Consequently~$\funrind(B)$
and~$\funrind(B')$ lie on the same branch~$B'$, and thus are comparable.
For the sake of contradiction, suppose~$\funrind(B')\sqsubset\funrind(B)$,
then by definition of~$\funvind$ we obtain~$\funvind(B)\notin B'$. Contradiction.
The remaining case is the expected $\funrind(B)\sqsubseteq\funrind(B')$.
\end{proof}
%%%%%%%%%%%%%%%%%%%%%%%%%%%%%% LEMMA %%%%%%%%%%%%%%%%%%%%%%%%%%%%%%
We are ready to establish the two properties wanted for~$\funvind$.
\begin{lemma} \label{lemma:vind-properties}
The following holds.
\begin{enumerate}[(1)]
\item For every node $x$, $\card{\funvind^{-1}(x)} < \kindex$.
\item If $\funvind(B') \sqsubset \funvind(B)$, then $\funvind(B) \notin B'$.
\end{enumerate}
\end{lemma}
%%%%%%%%%%%%%%%%%%%%%%%%%%%%%% LEMMA %%%%%%%%%%%%%%%%%%%%%%%%%%%%%%
\begin{proof}
(1): Let~$B$ be an infinite branch such that~$\funvind(B)=x$ and
let~$y = \funrind(B)$.  By Lemma~\ref{lemma:vind-rind}, important
branches with the same $\funvind$ also have the same $\funrind$ and
hence $\funvind^{-1}(x)\subseteq\funrind^{-1}(y)$.
The desired bound follows from Lemma~\ref{lemma:bounded-rind}.

(2): By Lemma~\ref{lemma:vind-rind}, if $\funvind(B') \sqsubset
\funvind(B)$ then $\funrind(B') \sqsubseteq \funrind(B)$.  If
$\funrind(B') \sqsubset \funrind(B)$, the claim follows by definition
of $\funvind(B)$. The case $\funrind(B') = \funrind(B)$ is not
possible since this would imply that $\funvind(B') = \funvind(B)$
because they are both lying on the branch $B$ according to the
assumption $\funvind(B') \sqsubset \funvind(B)$.
\end{proof}
Now, for $X \in \atoms$ such that $I(X)$ is an infinite branch we
define $\funbind(X)$ to be $\funvind(I(X))$. The distribution induced by
$\funbind$ is strongly sparse:
%%%%%%%%%%%%%%%%%%%%%%%%%%%%%% LEMMA %%%%%%%%%%%%%%%%%%%%%%%%%%%%%%
\begin{lemma}[strong sparsity]\label{lemma:branch-sparsity}
For $Z$ a finite zone of frontier $F$,
$\card{\funbind^{-1}(Z)} \le \kindex^2 \card{F}$.
\end{lemma}
%%%%%%%%%%%%%%%%%%%%%%%%%%%%%% LEMMA %%%%%%%%%%%%%%%%%%%%%%%%%%%%%%
\begin{proof}
Assume that $\funvind(B), \funvind(B') \in Z$ for two branches $B$ and
$B'$.  If the two branches exit $Z$ at the same point, i.e., if $B
\cap F = B' \cap F$, then $B$ and $B'$ do not differ inside $Z$. As
$\funvind(B), \funvind(B') \in Z$ we conclude that $\funvind(B) \in
B'$ and $\funvind(B') \in B$. Applying
Lemma~\ref{lemma:vind-properties}~(2) yields $\funvind(B) =
\funvind(B')$.

According to Lemmas~\ref{lemma:bounded-branchindex} and
\ref{lemma:vind-properties}~(1), the number of atoms $X$ that share
the same value $\funbind(X)$ is bounded by $\kindex^2$. This shows
that for each exit of the zone $Z$ there are at most $\kindex^2$
branches $B$ with $\funvind(B) \in Z$ leaving $Z$ through that
exit. Hence we obtain that $\card{\funbind^{-1}(Z)} \le \kindex^2
\card{F}$.
\end{proof}

%-------------------------------------------------------------------------------
\subsection{Car parking} \label{sub:car-parking}
%-------------------------------------------------------------------------------
In this subsection, our goal is to spread the indices around the tree
such that each index ends in exactly one position.  This can be seen
as parking vehicles.  In the beginning there are cars (indices) placed
in the nodes of the tree, possibly more than one at the same position,
and we aim at parking each of them in one node, i.e., attaching a
single node to each of those cars.  This is obviously not possible in
general but we shall prove that, under a sparsity constraint on the
distribution of vehicles, it is possible to attach a single parking
place to each car, and furthermore that the mapping that, given a car,
tells where to park it, can be described by a WMSO-formula.

For this purpose, we have to describe distributions and other kinds of
mappings that involve integers in their domain or image by
WMSO-formulas. These integers will always be bounded by some constant
$K$ and hence we can split the WMSO-definition into several formulas,
one for each number that may be involved.  Formally, we say that a
relation $R \subseteq \dom(t)^r \times I$ for some finite $I
\subseteq \mathbb{Z}$ is WMSO-definable if there are WMSO-formulas
$\phi_i(x_1, \ldots, x_r)$ for each $i \in I$ such that $(u_1,\ldots,
u_r, i) \in R$ iff $t,u_1, \ldots, u_r \models \phi_i(x_1, \ldots,
x_r)$. Note that a $K$-sparse distribution can be seen as a relation
of this kind since $0 \le D(x) \le 3K$ for every $x \in \dom(t)$.

\begin{definition}
Given a distribution~$D$, a \intro{placement}~$P$ for~$D$ is an
injective partial mapping from~$\dom(t)\times\nats$ to~$\dom(t)$ such
that~$P(x,i)$ is defined iff $i\in[D(x)]$.
\end{definition}

A flow, defined below, can be seen as a kind of instruction on how to
spread the values of a distribution to obtain a placement. In the
vehicles description, this is the number of cars which will have to
cross an edge in order to reach the final placement.

Recall that, for simplicity reasons, we assume that all the nodes of a
tree~$t$ have either $2$ successors or no successors, i.e., for all
nodes~$u$ we have $u0\in\mathit{dom}(t)$ iff $u1\in\mathit{dom}(t)$.
This assumption is not essential but simplifies the definitions and
allows to avoid case distinctions.
\begin{definition}
A \intro{flow} is a mapping~$f$ from the nodes of~$t$ to $\mathbb{Z}$.
A flow~$f$ is \intro{compatible} with a distribution~$D$
if for all inner nodes~$x$, $D(x) + f(x)\leq 1+f(x0)+f(x1)$ and
for every leaf~$x$, $D(x) + f(x) \leq 1$.
A flow~$f$ is \intro{bounded} by
a constant~$K$ if $\card{f(x)}\leq K$ for each node~$x$.
\end{definition}
In this definition, $f(x)$ is interpreted as the number of cars
crossing the edge from the ancestor of~$x$ to~$x$. In case of a
negative value, $-f(x)$ cars are driving from~$x$ to its ancestor.
The condition of $f$ being compatible with $D$ states that after
distributing all the cars according to the flow there is at most one
car remaining at each node. One should note here that, according to
our definition, it is possible that $f(\epsilon) < 0$. With the above
intuition this would mean that one has to send $-f(\epsilon)$ cars to
the (non-existing) ancestor of the root.  We need this case when
constructing flows on finite subtrees of a given infinite tree.

In the following we show that for a $K$-sparse distribution there is
a compatible flow that is bounded. From this flow we then compute a
placement for the distribution. We start by defining a flow on finite
trees (which can then also be used to deal with finite subtrees of a
given infinite tree).

%%%%%%%%%%%%%%%%%%%%%%%%%%%%%% LEMMA %%%%%%%%%%%%%%%%%%%%%%%%%%%%%%
\begin{lemma}\label{lemma:flow-finite}
For every finite tree~$t$ and every WMSO-definable $K$-sparse
distribution~$D$ over~$t$, there exists a WMSO-definable flow $f$ that
is compatible with $D$, bounded by $2K+1$, and such that for each node
$x$ there is a zone $Z_x$ rooted in $x$ of frontier $F_x$ with
\begin{enumerate}[(1)]
\item $D(Z_x)+f(x)=\card{Z_x}+K(\card{F_x}-1)$,
\item and $f(y)\geq K$ for all $y\in F_x$ different from~$x$.
\end{enumerate}
\end{lemma}
%%%%%%%%%%%%%%%%%%%%%%%%%%%%%% LEMMA %%%%%%%%%%%%%%%%%%%%%%%%%%%%%%
\begin{proof}
First note that (1) implies $f(x) \ge -K$ for each $x$ since $D$ is
$K$-sparse.  We define the values $f(x)$ and the zones $Z_x$
inductively starting at the leaves. These definitions directly imply
that $f$ is compatible with $D$.

The base case of a leaf~$x$ is straightforward: we set~$f(x)$ to
be~$1-D(x)$ and~$Z_x=\{x\}$.  By the hypothesis of sparsity, we have
that~$D(x)\leq K+1$ and hence $|f(x)|$ is bounded by $2K+1$.

Let $x$ be an inner node and assume that the values $f(x0)$, $f(x1)$
and the zones $Z_{x0}$, $Z_{x1}$ are already defined. We set $f'(x0) =
\min(K,f(x0))$ and $f'(x1) = \min(K,f(x1))$.  Let us now define $f(x)$
to be $1+f'(x0)+f'(x1)-D(x)$ and~$Z_x$ to contain the node $x$, the
nodes of $Z_{x0}$ if $f(x0)<K$, and the nodes of $Z_{x1}$
if~$f(x1)<K$.  By the hypothesis of induction, we indeed obtain that
$D(Z_x)+f(x)=\card{Z_x}+K(\card{F_x}-1)$.  We illustrate this only for
the case $f'(x0) < K$ and $f'(x1) = K$, the other cases are
similar. In this case $Z_x = Z_{x0} \cup \{x\}$, $\card{F_x} =
\card{F_{x0}} + 1$, and $f(x) = 1 + f(x0) + K -D(x)$. From this we get
the following sequence of equalities:
\begin{eqnarray*}
 \\
D(Z_x) + f(x) & = & D(Z_{x0}) + D(x) + 1 + f(x0) + K - D(x) \\
&=&  D(Z_{x0}) + f(x0) + 1 + K \\
&=&  \card{Z_{x0}} + K(\card{F_{x0}} - 1) + 1 + K \\
&=&  \card{Z_{x0}} + 1 + K\card{F_{x0}} \\
&=&  \card{Z_{x}} + K(\card{F_{x}} - 1).
\end{eqnarray*}
From the definition of $f(x)$ it is clear that $f(x)\leq 2K+1$.  As
mentioned before, $f(x) \ge -K$ and thus $|f(x)|$ is bounded by
$2K+1$.

It is obvious that this flow, which has only a bounded number
of possible values and is defined inductively, is WMSO-definable. This
definition can be done by requiring the existence of sets $X_{-K},
\ldots, X_{2K+1}$ such that a node $x$ is in $X_i$ iff $f(x) = i$.
This can be directly expressed if $x$ is a leaf. Otherwise it is a
simple statement on the membership of the successors $x0$ and $x1$
of $x$.
\end{proof}

%%%%%%%%%%%%%%%%%%%%%%%%%%%%%% LEMMA %%%%%%%%%%%%%%%%%%%%%%%%%%%%%%
\begin{lemma}\label{lemma:flow-infinite}
For every infinite tree~$t$ and every WMSO-definable $K$-sparse
distribution~$D$, there exists a WMSO-definable flow $f$ bounded
by~$7K$ that is compatible with~$D$ such that $f(\epsilon) = 0$.
\end{lemma}
%%%%%%%%%%%%%%%%%%%%%%%%%%%%%% LEMMA %%%%%%%%%%%%%%%%%%%%%%%%%%%%%%
\begin{proof}
As a $K$-sparse distribution restricted to a finite subtree of $t$ is
also $K$-sparse on this subtree, we can apply
Lemma~\ref{lemma:flow-finite} to define the values of $f(x)$ for the
nodes $x$ that are not on infinite branches of $t$.

Let~$B$ be the set of nodes appearing in some infinite branch. We
define inductively for any node~$x\in B$ a flow~$f(x)$ such that $0
\le f(x) \le 7K$. We only consider non-negative values since on
infinite branches we never reach a leaf and hence there is no need for
an upward flow.

We start by setting $f(\epsilon) = 0$.  For $x \not= \epsilon$
let~$y\in B$ be the father of~$x$.  Three cases may happen. If at node
$y$ only one of its children is in $B$ (case (a)), we forward
everything to this node. Otherwise (cases (b) and (c)), we forward at
most $5K$ to the left child and the rest to the right child. The
formal definitions are given below, where the the $\max$ operator is
only used to avoid negative flows.
\begin{enumerate}[(a)]
\item If $x$ is the only child of~$y$ in~$B$, then we set
 \[
f(x) = \max(0,f(y)+D(y)-f(x')-1)
\]
where~$x'$ is the other child of~$y$.
\item If the two children of~$y$ are in~$B$ and~$x$ is the left child,
	then we set
\[
f(x) = \max(0,\min(5K,f(y)+D(y)-1)).
\]
\item If the two children of~$y$ are in~$B$ and~$x$ is the right
	child, then we set
\[
f(x)=\max(0,f(y)+D(y)-1-5K).
\]
\end{enumerate}
Obviously, $f(x) \ge 0$ in all cases.  We show that if $x$ is of type
(b) or (c), then $f(x) \le 5K$, and if $x$ is of type (a), then $f(x)
\le 7K$.

In case (b), $f(x) \le 5K$ follows directly from the definition and in
case (c) from $f(y) \le 7K$ (by induction) and $D(y) \le 3K+1$ ($D$ is
$K$-sparse).

For $x$ as in (a) we cannot use a local argument but we have to go
upwards until we reach a node that has a flow of at most $5K$. Such a
node must exist because we eventually meet a node of type (b) or (c),
or the root, which has flow $0$. All nodes we met before must be of
type (a).

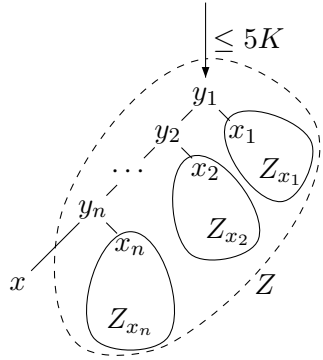
\begin{figure}
\begin{picture}(40,50)(0,-10)
\gasset{Nframe=n,AHnb=0,Nw=5,Nh=5}
\node(x)(0,0){$x$}
\node(yn)(10,10){$y_n$}
\node(xn)(15,5){$x_n$}
\node(dots)(15,15){$\cdots$}
\node(y2)(20,20){$y_2$}
\node(x2)(25,15){$x_2$}
\node(y1)(25,25){$y_1$}
\node(x1)(30,20){$x_1$}
\node(top)(25,40){}

\drawedge(yn,x){}
\drawedge(yn,xn){}
\drawedge(dots,yn){}
\drawedge(y2,dots){}
\drawedge(y2,x2){}
\drawedge(y1,x1){}
\drawedge(y1,y2){}
\drawedge[AHnb=1](top,y1){$\le 5K$}

\node(zn)(15,-5){$Z_{x_n}$}
\node(z2)(28,7){$Z_{x_2}$}
\node(z1)(35,15){$Z_{x_1}$}
\node(z)(33,0){$Z$}
\gasset{Nframe=y}
\drawccurve(15,7)(10,-7)(20,-7)
\drawccurve(23,17)(23,4)(32,6)
\drawccurve(28,22)(35,11)(39,19)
% Zone Z
\drawccurve[dash={1 1}](25,30)(6,10)(8,-8)(20,-8)(40,20)
\end{picture}
\caption{Proof of Lemma~\ref{lemma:flow-infinite}: The flow in nodes of
type (a) is bounded by $7K$.} \label{fig:typea}
\end{figure}

The following definitions are illustrated in Figure~\ref{fig:typea}.
The choice of the $y_i$ being the left successors in the figure is
arbitrary and only for matters of presentation. Let $y_n, \ldots, y_1$
be such that $y_n$ is the father of $x$, $y_{i-1}$ is the father of
$y_i$ for all $i \in \{2, \ldots, n\}$, $f(y_1) \le 5K$, and $f(y_2),
\ldots, f(y_n) > 5K$. As mentioned above, $y_2, \ldots, y_n$ are of
type (a).

Let $x_1, \ldots, x_n$ be such that $y_i$ is the father of $x_i$, $x_n
\not = x$, and $x_i \not= y_{i+1}$ for $i \in \{1, \ldots, n-1\}$,
i.e., $x_i$ is the brother of $y_{i+1}$. For the $x_i$ the flow is
defined using Lemma~\ref{lemma:flow-finite} because they are not lying
on an infinite branch. Hence there are zones $Z_{x_i}$ rooted at $x_i$
of frontier $F_{x_i}$ such that
\begin{equation}
D(Z_{x_i}) + f(x_i) = \card{Z_{x_i}} + K(\card{F_{x_i}} - 1). \label{eq:zoneZxi}
\end{equation}
Let $Z = \bigcup_{i=1}^n (\{y_i\} \cup Z_{x_i})$ and let $F$ be the
frontier of $Z$. Then
\begin{eqnarray}
\card{Z} &=& n + \sum_{i=1}^n \card{Z_{x_i}} \label{eq:cardZ} \\
\card{F} &=& 2 + \sum_{i=1}^n (\card{F_{x_i}}-1) \label{eq:cardF} \\
\sum_{i=1}^n D(y_i) &=& D(Z) - \sum_{i=1}^n D(Z_{x_i})  \label{eq:Dyi}
\end{eqnarray}
Since $y_2, \ldots, y_n$ are of type (a) and furthermore their flow is
bigger than $5K$ (and hence bigger than $0$), we get
\[
f(x) = f(y_1) + \sum_{i=1}^n (D(y_i) - 1 - f(x_i)).
\]
We know that $f(y_1) \le 5K$ and hence it remains to be shown that
$\sum_{i=1}^n (D(y_i) - 1 - f(x_i)) \le 2K$. This can be deduced as
follows:
\begin{eqnarray*}
\sum_{i=1}^n (D(y_i) - 1 - f(x_i))
 &\stackrel{(\ref{eq:Dyi})}{=} &
   D(Z) - n - \sum_{i=1}^n (D(Z_{x_i}) + f(x_i))
\\
& \stackrel{(\ref{eq:zoneZxi})}{=} &
   D(Z) - n - \sum_{i=1}^n (\card{Z_{x_i}} + K(\card{F_{x_i}} - 1))
\\
& \stackrel{(\ref{eq:cardF})}{=} &
   D(Z) - n - K(\card{F} - 2) - \sum_{i=1}^n \card{Z_{x_i}}
\\
& \stackrel{\mbox{{\scriptsize $K$-sparse}}}{\le} &
   \card{Z} + K\card{F} - n -  K(\card{F} - 2) - \sum_{i=1}^n
   \card{Z_{x_i}}
\\
&\stackrel{(\ref{eq:cardZ})}{=} & 2K
\end{eqnarray*}
That this flow $f$ is compatible with $D$ and that it is
WMSO-definable can easily be deduced from the definitions.
\end{proof}

We are now ready to establish our placement Lemma.
%%%%%%%%%%%%%%%%%%%%%%%%%%%%%% LEMMA %%%%%%%%%%%%%%%%%%%%%%%%%%%%%%
\begin{lemma}[car parking] \label{lemma:car-parking}
For every tree~$t$ and every WMSO-definable $K$-sparse distribution~$D$,
there exists a WMSO-definable placement for~$D$. If $t$ is finite, we
additionally require that $D(\dom(t)) \le \card{\dom(t)}$.
\end{lemma}
%%%%%%%%%%%%%%%%%%%%%%%%%%%%%% LEMMA %%%%%%%%%%%%%%%%%%%%%%%%%%%%%%
\begin{proof}
According to Lemmas~\ref{lemma:flow-finite} and
\ref{lemma:flow-infinite} we know that there is a WMSO-definable flow
$f$ that is compatible with $D$. For simplicity, we first assume that
$f(\epsilon) = 0$, which is always the case for infinite trees
(Lemma~\ref{lemma:flow-infinite}). If $f(\epsilon) > 0$ for finite
trees, then we can simply redefine $f(\epsilon) = 0$ without changing
the property of $f$ being compatible with $D$. If $t$ is finite and
$f(\epsilon) < 0$, then we cannot simply set $f(\varepsilon) = 0$
because this would affect the compatibility of $f$ with $D$.  At the
end of the proof we briefly explain how to treat this case.

The general strategy for defining the placement is the following:
\begin{itemize}
\item From each node we send all the cars except one to its
  neighbors. The number of cars sent to each neighbor is described
  by the flow.
\item If we follow this strategy, then each edge in $t$ is crossed by
  the cars in only one direction. Hence, a car cannot visit the same
  node twice. This means that it might be sent up in the tree for some
  steps and from some point onwards it is only sent downwards.
\item To be sure that each car will be parked after finite time we
  order all the cars that cross a node (as described by $D$ and $f$)
  according to a fixed strategy and we also fix a scheme for
  distributing the cars to the neighboring nodes.
\item This ordering will ensure that the index of a car decreases each
  time it is sent down in the tree. As described above, each car is
  sent up in the tree only a finite number of times. Hence, if we
  always park the car that is first in the ordering at a specific
  node, then each car will eventually be parked.
\end{itemize}
To show that this strategy can be realized by WMSO-formulas we first
describe the ordering of cars that we use and then define formulas

\begin{itemize}
\item $\send_i(x,y)$ meaning that the $i$th car at node $x$ is sent to
  node $y$.
\item $\drive_{i,j}(x,y)$ meaning that the $i$th car at node $x$ is
  sent to $y$ and is car number $j$ at $y$.
\item $\start_i(x)$ meaning that the $i$th car at node $x$ does not
  come from another node.
\item $\itinerary_i(x,X_1, \ldots, X_K, y)$ meaning that the $i$th car
  at node $x$ will be parked at node $y$ using an itinerary that is
  described by the sets $X_1, \ldots, X_K$.
\end{itemize}
To define these formulas we first have to introduce some notation. To
avoid case distinctions we define $f^+(x) = \max(f(x),0)$ and $f^-(x)
= \min(f(x),0)$. Furthermore, we assume by convention that for a leaf
$x$ the values $f^+(x0)$, $f^+(x1)$, $f^-(x0)$, $f^-(x1)$ are defined,
and are all set to $0$.

Then the number of cars crossing a node $x$ is $f^+(x) + f^-(x0) +
f^-(x1) + D(x)$.  Since all the values involved in this expression are
bounded and since we can increase $K$ without affecting the
$K$-sparsity of $D$, we can assume that $f^+(x) + f^-(x0) +
f^-(x1) + D(x) \le K$ for all $x$.

Since $f$ is WMSO-definable, we can also assume that there are formulas
$\phi_i^+(x)$ and $\phi_i^-(x)$ defining $f^+$ and $f^-$. Then
expressions of the form $i_1 + f^+(x) + f^-(y) = i_2$ for $i_1,i_2 \in
[K]$ can easily be expressed as Boolean combinations of the formulas
$\phi_i^+$ and $\phi_i^-$. The use of expressions of this kind
simplifies the presentation of the formulas.

We start by giving the orderings used in the definitions of the
formulas $\send_i$ and $\drive_{i,j}$. The cars that cross a node $x$
will be distributed in the following order that we refer to as the
distribution order:
\[
\begin{array}{|c|c|c|c|}
\hline
f^+(x) & f^-(x0) & f^-(x1) & D(x) \\
\hline
\end{array}
\]
That is, we first distribute the cars that come from the father of $x$, then
the cars that come from the left child of $x$, and so on. It remains
to fix where to send the cars.
\[
\begin{array}{|c|c|c|c|}
\hline
1 & ~f^-(x)~ & ~f^+(x0)~ & ~f^+(x1)~ \\
\hline
\end{array}
\]
This means that the first car in the distribution order is parked
in the node $x$. The next cars are sent to the father of $x$ if $f(x)$ is
negative. The following cars in the distribution order are send to the
left child of $x$ and the remaining cars to the right child of
$x$.

To illustrate this, consider the following example with $x = y0$,
$f^+(y) = 3$, $f^-(x) = 5$, and $f^+(y1) = 7$.
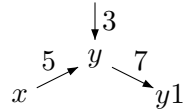
\begin{center}
\begin{picture}(20,20)
\gasset{Nframe=n,Nw=5,Nh=5}
\node(y)(10,10){$y$}
\node(y0)(0,5){$x$}
\node(y1)(20,5){$y1$}
\node(py)(10,20){}

\drawedge(y,y1){$7$}
\drawedge(y0,y){$5$}
\drawedge(py,y){$3$}
\end{picture}
\end{center}
Let us first see how
the cars at $y$ are ordered according to the distribution order. The
first three are the ones coming from the father of $y$. The next $5$
are those coming from $x$. There are no cars coming from $y1$, and the
last cars are those from $D(y)$.  Now, we would like to know what
happens to the $4$th car at $x$. The first car at $x$ stays at
$x$. The next $5$ cars are sent to $y$, that is, the fourth car at $x$
is the third car sent from $x$ to $y$. According to the distribution
order at $y$ described before, this car becomes car number $6$ at $y$.

This is expressed by the following formulas, where the first two
are only defined for $2 \le i \le K$ because the first car is always
kept at the current position.
\begin{itemize}
\item For $2 \le i \le K$:
\[
\send_i(x,y) :=
\begin{array}[t]{l}
\left[(x = y0 \, \lor \, x = y1) \, \land \, 1 < i \le C_1\right] \\
{}\lor
[x0 = y \,  \land \, C_1 < i \le C_2] \\
{}\lor
[x1 = y \,  \land \, C_2 < i \le C_3] \\
\end{array}
\]

Here $C_1 = 1+ f^-(x)$, $C_2 = 1 + f^-(x) + f^+(x0)$, and $C_3 = 1 +
f^-(x) + f^+(x0) + f^+(x1)$.
\item For $2 \le i \le K$:
\[
\drive_{i,j}(x,y) :=
\begin{array}[t]{l}
\send_i(x,y) \land ~\\
\big(\left[x = y0  \, \land \, j = i-1 + f^+(y) \right] \\
{}\lor
[x = y1 \, \land \, j = i-1 + f^+(y) + f^-(y0)] \\
{}\lor
[x0 = y \,  \land \, j = i - 1 - f^-(x)] \\
{}\lor
[x1 = y \,  \land \, j = i - 1 - f^-(x) - f^+(x0) ]\big)
\end{array}
\]
\item For $1 \le i \le K$:
\[
\start_i(x) :=
f^+(x) +f^-(x0) + f^-(x1) < i \le f^+(x) +f^-(x0) + f^-(x1) + D(x)
\]
\item For $1 \le i \le K$:
\[
\itinerary_i(x,X_1, \ldots, X_K, y) :=
\begin{array}[t]{l}
\mathrm{disjoint(X_1, \ldots, X_K)} \land ~ \\
x \in X_i \land \start_i(x) \land X_1 = \{y\} \\
{}\land \bigwedge\limits_{j=2}^K (\forall z \in X_j \bigvee\limits_{j' \in [K]} \exists
z' \in X_{j'} \;:\; \drive_{j,j'}(z,z'))
\end{array}
\]
This formula states that the $i$th car at $x$ starts there.  The free
set variables describe the set of positions that this car crosses,
where a position is included in $X_m$ if the car is the $m$th one at
this position. Finally, it states that $y$ is the only position where
the car is first in the ordering. Hence it will stop at $y$.
\end{itemize}
Then the formulas $\psi_i(x,y)$ defining the placement are given by
\[
\psi_i(x,y) = \exists X_1, \ldots X_K (\itinerary_i(x,X_1, \ldots, X_K,y)).
\]
As mentioned at the beginning of the proof we now discuss how to treat
the case $f(\epsilon) < 0$. Recall that this may only happen for
finite trees. In general, simply redefining $f(\epsilon) = 0$ may lead
to a flow that is not compatible with $D$ anymore. Therefore, we have
to use a different strategy.

The strategy for distributing the cars described above would lead to
$f^-(\epsilon)$ cars that ``get stuck'' at the root because, following the
flow, they should be sent upwards, and this is not possible.
This means that there
are at most $K$ cars (for simplicity assume exactly $K$ cars) that
start at some node but never arrive at some destination, i.e., there
are nodes $x_1, \ldots, x_K$ and $i_1, \ldots, i_K \in [K]$ such that
\[
\start_{i_j}(x_j) \land \lnot \exists y (\psi_{i_j}(x_{j},y))
\]
is satisfied. From the assumption $D(\dom(t)) \le \card{\dom(t)}$ we
can conclude that there remain at least~$K$ nodes where no car is
parked, i.e., $K$ nodes which are not image of the function defined by
the $\psi_i$'s.  Let $y_1, \ldots, y_K$ be the~$K$ first such nodes
for some WMSO-definable order (this is possible since the tree is
finite).  We can now extend this function by ordering the $x_1,
\ldots, x_K$ and map the $i_j$th car from $x_j$ to $y_j$.  Note that
these definitions are expressible in WMSO.  In this way, we obtain a
modification of the function defined by the $\psi_i$'s into a
placement for $D$.
\end{proof}

%-------------------------------------------------------------------------------
\subsection{Proof of Theorem~\ref{tHeorem:main-result}} \label{sub:mainproof}
%-------------------------------------------------------------------------------
We can now prove Theorem~\ref{tHeorem:main-result} as stated in
Section~\ref{section:main} by combining the previous results.

For $X \in \atoms$ let
\[
\funind(X) = \left\{
\begin{array}{l}
\funbind(X) \mbox{ if $I(X)$ is an infinite branch,} \\
\funsind(X) \mbox{ otherwise.}
\end{array} \right.
\]
We construct a formula $\phiind(X,y)$ that associates to each $X
\in \atoms$ its index $y = \funind(X)$.  From the definitions of
$I(X)$, $\funsind(X)$, and $\funbind(X)$ it is clear that we can
construct such a formula. Note that in the definition of this formula,
we do not have to explicitly represent infinite sets (though $I(X)$
may be infinite) because we can construct a WMSO formula
$\phi_{\mathit{imp}}(X,x)$ that associates to $X \in \atoms$ its
important nodes. From this one can construct WMSO definitions of
$\funsind(X)$ and $\funbind(X)$, and hence the formula $\phiind(X,y)$
is also WMSO.

Then, we compute the distribution $D$ defined by $D(x) =
\card{\funind^{-1}(x)}$. Since this distribution is $(\ksparse +
\kindex^2)$-sparse by Lemmas~\ref{lemma:sparsity}
and~\ref{lemma:branch-sparsity}, it is also WMSO-definable using the
formula $\phiind$. Let $K$ be a constant such that $D(x) \le K$ for
all nodes $x$.  Applying Lemma~\ref{lemma:car-parking}, we obtain a
WMSO-definable placement $P$ for $D$. One should note that for finite
$t$ the assumption $D(\dom(t)) \le \card{\dom(t)}$ is satisfied
because $D(\dom(t))$ is the number of elements in the set $E$, and
$\interpretation(t)$ being isomorphic to $\powerF(E)$ implies that $t$
has at least as many elements as $E$.

Let $\psi_i(x_1,x_2)$ for $i \in [K]$ be the formulas defining $P$.
Now, we order all the $X \in \atoms$ with the same index. A
possible definition for such an ordering is $X < Y$ if the
lexicographically smallest node that is not in $X \cap Y$ is in
$X$. Then one can construct WMSO-formulas $\theta_i(X)$ stating that
$X$ is the $i$th set in the ordering among those that have the same
index as $X$.

The WMSO-formula $\code(X,x)$ that attaches $X$ to its final
position $x$ is then defined as follows:
\[
\code(X,x) = \bigvee_{i \in [K]}(\theta_i(X) \land \exists y (\phiind(X,y)
\land \psi_i(y,x)).
\]
\qed

One should note here that without the results from
Subsection~\ref{sub:car-parking} we can obtain a weaker version of
Corollary~\ref{corollary:main-general} by replacing the interpretation
$\interpretation_2$ by a transduction, i.e., an interpretation that
can use a fixed number $K$ of copies of the given structure. Such a
transduction can be realized using the formula $\phiind$ instead of
$\code$.  In particular one could use this weaker version of the
result in all the applications, but at the price of some notational
and technical overheads.

From this point of view, Lemma~\ref{lemma:car-parking} can also be
seen as a result on the question under which conditions a
(W)MSO-transduction is equivalent to a (W)MSO-interpretation on binary
trees. Namely, if the distribution defined by the transduction, i.e.,
the function assigning to each node the number of times it is used in
the result of the transduction, is $K$-sparse (for some $K$) on each
tree. And this presentation can be used either for WMSO-transductions or
MSO-transductions.

\subsection*{Acknowledgments}\label{section:ack}
We are particularly grateful to Vince B\'{a}r\'{a}ny for his
contribution to the random-graph proof, as well as his comments on
previous versions of this work. We also thank Achim Blumensath for
commenting on earlier stages of this work, and the anonymous referees
for their helpful comments.

\newcommand{\etalchar}[1]{$^{#1}$}

\end{document}